\documentclass[11pt, oneside, reqno]{amsart}   	
\usepackage{geometry}                		
\usepackage{amssymb}
\usepackage{mathrsfs}
\begin{document}
\begin{center}
Non-Ultralocality and Causality in the Relational Framework of Canonical Quantum Gravity
\\ 
\vspace{2ex}
P. G. N. de Vegvar \\ 
\vspace{2ex}
SWK Research\\
1438 Chuckanut Crest Dr., Bellingham, WA 98229, USA\\
  \vspace{2ex}
Paul.deVegvar@post.harvard.edu\\ 
\vspace{2ex}
PACS: 04.20.Gz, 04.20.Fy  
\end{center}
The relational framework of canonical quantum gravity with non-ultralocal constraints is explored. After demonstrating the absence of anomalies, a spatially discretized version of the relational framework is introduced. This allows the application of Lieb-Robinson bounds to on-shell monotonic gauge-flow when there is a continuous external ``time" parameter. An explicit Lieb-Robinson bound is derived for the differential on-shell evolution of the operator norm of the commutator of discretized Dirac observables, demonstrating how a local light cone-like causal structure emerges. Ultralocal constraints do not permit such a structure to arise via Lieb-Robinson bounds. Gauge and (3+1)-diffeomorphism invariance of  the light-cone is discussed along with the issues of quantum fluctuations, the nature of the nonlocalities, the spatial continuum limit, and possible links to non-commutative geometry.
\vspace{2.8in}
\pagebreak

\section{Introduction and Motivation}

Locality and causality have played pivotal roles in gravitational physics for nearly a century. 
While the concept and important consequences of local light cones are well established in general relativity, causality in canonical loop quantum gravity and in spin
foam models has proven more difficult to elucidate. Almost 20 years ago Smolin \cite{Smolin}
pointed out the importance of the emergence of classical long range correlations for non-pertubative theories of quantum 
gravity in some appropriate semiclassical limit. These correlations are then reflected in how one thinks about causality in quantum field theory on a fixed classical background spacetime,  
specifically that local observables at space-like separation commute (micro-causality). 
During the intervening years, there have been several studies of causality in the background independent 
fully quantum regime, particularly utilizing spin foam 
(covariant) models. However, generally the spin foam models do not support causal correlations, unless they are either assumed at the outset \cite{Marko_Smolin}\cite{Livine_Terno} or
involve some kind of alteration of the vertex amplitude intended to describe a local orientation \cite{Oriti_Livine}.
On the other hand, 
approaching the issue of semiclassical causality from the canonical point of view is even more conceptually challenging since that approach is a ``time-less" formalism. Consequently, the 
consensus expectation is that micro-causality will emerge from some as yet to be developed semiclassical limit of quantum gravity, however to date such a 
causal limit  for quantum gravity remains lacking. Thus it is 
mysterious  that micro-causality occupies such a foundational place in general relativity, quantum field theory, and the standard model of particle physics, yet is still so elusive from a 
background independent quantum gravity point of view. 
In a broader context, the semiclassical regime of quantum gravity is of importance not only from the perspective of causal correlations, but also as a  general testing ground to examine whether a theory of quantum gravity  can behave in familiar classical ways in some suitable limit. \\

Here we take a simple first step towards understanding these questions starting from an unexpected direction. We adopt the relational framework approach to canonical quantum gravity 
which has been developed over several decades
\cite{Kuchar}\cite{Rovelli}\cite{Dittrich_RF}\cite{Dittrich_GR}\cite{Thiemann_RF}.
Then we explore the case where the constraints are non-ultralocal, and for reasons to be discussed, limit our considerations to the on-shell physics. 
Next, we apply Lieb-Robinson bounds, originally introduced in the 1970s to describe solid-state spin systems \cite{Lieb}, to a spatially 
discretized version of the relational framework and demonstrate how a suitably gauge invariant 
differential local light cone for discretized Dirac observables may be constructed. In essence, the local light-cone emerges from on-shell non-ultralocality of 
constraints in a quite general sense within spatially 
discrete relational framework models with smooth monotonic gauge flow described by an external ``time" parameter. 
If the constraints are taken to be ultralocal, then this Lieb-Robinson-based causal structure collapses. Quantum fluctuations act to disrupt 
the local light-cone structure, and a set of general criteria are presented which are sufficient for the Lieb-Robinson-based local light-cone structure to survive the quantum-classical tug-of-war.  \\

The outline of the remainder of the paper is as follows: The relational framework is briefly recapitulated in section 2. Section 3 studies non-ultralocality, its freedom from anomalies, 
and introduces ``patchy" gauge flow. It concludes with a 
description of the spatially discretized model that is used later on. Section 4 provides an introduction to two non-relativistic versions of earlier Lieb-Robinson bounds from the literature: The first for Heisenberg 
operator evolution via time independent Hamiltonians, and another
more mathematically sophisticated approach for time-dependent Hamiltonians, both on general lattices or networks. In section 5 a relativistic differential-time expression for the 
Lieb-Robinson local light-cone is 
derived for the case of an external time parameter acting as the synchronizing conductor of the relational framework's clock variable symphony orchestra. 
Section 6 discusses gauge invariance and other 
properties of the relational local light-cone. 
Through a series of questions and answers, Section 7 examines issues related to the continuum limit of the spatial discretization, the nature of nonlocality necessary for the 
Lieb-Robinson local light-cone, a possible link to non-commutative geometry, and the role of quantum fluctuations. The paper concludes with a brief summary and self-criticism of the model in section 8. \\

\section{Review of the Relational Framework}

In this section we briefly review the necessary points of the relational framework formalism. For further details please see \cite{Dittrich_RF}\cite{Dittrich_GR}\cite{Han}.
 The essential idea behind the relational framework is to construct Dirac (gauge invariant)  
observables from gauge variant (partial) observables. This all starts from the classical phase space $\mathscr {M}$ description of a re-parametrization invariant system whose dynamics is 
described by a (canonical) Hamiltonian that consists entirely of a linear combination of constraints. 
In the relational framework, quantization occurs on the reduced phase space. We start by describing the classical formalism.
Consider then a set of first-class constraints ${C_I}$ with $I\in \mathscr{I}$, an arbitrary index set. For the case of canonical 4-dimensional general relativity, the index $I$ includes both a continuous 3-coordinate index $y(I)$, labeling a point $\sigma$ on the 3-dimensional manifold $\Sigma$, as well as a discrete index $i(I)$. The latter index ranges from 0 through $N_{c} -1$, and labels the $N_c$ first-class constraints (gauge conditions) at each point $\sigma$. 
These constraints satisfy the first-class condition $\{C_I , C_J \} =f^K_{IJ}\, C_K$, where generally
 $f^K_{IJ}$ may be a structure function, possibly depending on phase space functions. We have assumed all second-class constraints have previously been handled by taking $\mathscr {M}$
 to be the surface in phase space where they all vanish, and that the Poisson bracket used above is the Dirac bracket. Next select a set of gauge variant phase space 
 functions $T^I$,  $I\in \mathscr{I}$ called clock functions or clock variables that coordinatize the gauge orbit of any point in phase space within a neighborhood of the (classical) constraint 
 surface (shell) $\bar{\mathscr M} \doteq \{ m \in \mathscr{M} | C_I(m)=0, \forall I\in\mathscr{I} \}$. The $T^I$ might include matter (non-gravitational) degrees of freedom. Then $A^J_I\doteq \{ C_I ,
 T^J \}$ is locally non-singular as an $(I, J)$ ``matrix," and one can define the transformed equivalent first-class constraints
 
\begin{equation}
C' _I \doteq \sum _{J} [A^{-1}]^J _I C_{J}.
\end{equation}
These obey $\{ C' _I , T^J \} \approx \delta ^J _I $, where $\approx $ denotes weak equality, that is equality on-shell, and the Hamiltonian vector fields $X_{C' _I} \doteq X_I$
weakly commute (i.e., commute on-shell). Gauge transformations for any phase space function $f$ and set $\beta ^I$ of reals can be written as
\begin{align}
\alpha _{\beta } (f) & \doteq \exp (X_{\beta} ) \cdot f = \sum ^{\infty}_{n=0} \frac{1}{n!}  (X_{\beta})^{n} \cdot f \\
X_{\beta} & \doteq \sum _{I}  \beta ^{I} X_I .   \nonumber 
\end{align}
If one is given a set of real-valued phase space constants (clock parameters) $\tau ^I$, a weakly gauge invariant (Dirac) observable associated with partial (gauge variant) observables $f$
and $T^I$ is
\begin{equation}
\mathscr{O} _{f} (\tau) \doteq \mathscr{O} [f] (\tau) \doteq [\alpha _{\beta} (f)]_{\alpha _{\beta} (T^I) = \tau ^I }.
\end{equation}
The motivating idea is that $\mathscr{O} [f] (\tau)$ represents the value of $f$ when the clock variables $T^{I}$ take the values $\tau ^{I}$; i.e., it is a gauge 
slice or fixing. $\mathscr{O} [f] (\tau)$ is a phase 
space function, and one must compute it first treating $\beta ^I$ as phase space constants, and only subsequently set $\beta ^{I} = \tau ^{I} - T^{I}$. \\

One also finds $\alpha_{\beta} (T^I) \approx  T^{I} + \beta ^{I}$ and $\mathscr{O}[T^{I}](\tau) \doteq \alpha_{\beta}(T^{I}) _{\alpha_{\beta}(T^{I})) = \tau^{I} } \approx \tau^{I}$ on-shell.
Also on the constraint surface $\mathscr{O} [f] (\tau)$ may be formally expanded as
\begin{equation}
\mathscr{O} [f] (\tau) \approx \sum ^{\infty} _{\{k_I =0\} } \Bigg( \prod _{I} \frac {(\tau ^{I} - T^{I})^{k_{I} }} { k_{I}!} \Bigg) \Bigg( \prod _{I} (X_{I})^{k_{I} } \Bigg) \cdot f.  \label {SeriesExpsn}
\end{equation}
One can also derive
\begin{align}
\mathscr{O} [f] (\tau) + \mathscr{O} [f'] (\tau) & = \mathscr{O} [f + f'] (\tau) \label {AddHomomorph}  \\
\mathscr{O} [f] (\tau) \:  \mathscr{O} [f'] (\tau) &  \approx  \mathscr{O} [f f'] (\tau) \label {MultHomomorph}  \\ 
\big\{\mathscr{O}[f](\tau ), \mathscr{O}[f'] (\tau) \big\} & \approx \big\{\mathscr{O}[f] (\tau ), \mathscr{O}[f'](\tau) \big\} _{D}  \approx \mathscr{O}\big[\{ f,f' \}_{D}\big] (\tau ),  \label{PBHomomorph} 
\end{align}
where the Dirac bracket is defined as
\begin{equation}
\{f,f'\}_{D} \doteq \{f,f'\} - \{f, C_{I}\} [A^{-1}]^{I}_{J} \{T^{J}, f' \} + \{f', C_{I}\} [A^{-1}]^{I}_{J} \{T^{J},f\}.
\end{equation}

The formalism simplifies considerably if one can choose canonical coordinates so that the clock variables $T^I$ are themselves some canonical coordinates. Then one has a complete set of 
canonical pairs partitioned as $(q^{a}, p_{a})$ and $(T^{I}, P_{I})$, where the $P_I$ are the canonical momenta conjugate to the $T^I$. Hence in a local neighborhood of the constraint surface 
one can write the constraints as the equivalent set
\begin{equation}
 \tilde {C}_{I} = P_{I} + h_{I}(q^{a}, p_{a}, T^{J}) \approx 0,
\end{equation}
and setting $P_{I} =-h_{I}(q^{a}, p_{a}, T^{J})$ formally solves the constraints.
One can also show that the canonical Dirac observables $Q^{A}(\tau ) \doteq \mathscr{O} [q^{a}] (\tau )$ and $P_{a} (\tau ) \doteq \mathscr{O} [p_a] (\tau )$ satisfy equal-$\tau$ canonical 
Poisson bracket relations. As discussed in \cite{Dittrich_RF}\cite{Han}, the $\tilde {C}_{I}$ comprise a strongly Abelian constraint algebra and obey 
$ C'_{I} = \tilde {C} _{I} \,+ \,\mathcal{O} (C^{2})$,
hence the Hamiltonian vector fields $X_{I}$ of $C' _{I}$ and $X_{\tilde{I}}$ of $\tilde{C}_{I}$  weakly commute. The relations (\ref{AddHomomorph}), (\ref{MultHomomorph}) and 
(\ref{PBHomomorph}) may be summarized by saying that $\mathscr{O}$ induces a weak algebra homomorphism w.r.t. pointwise addition and multiplication as well as a weak Dirac algebra 
homomorphism on $\{ f,f'\}_{D}$.  When neither $f$  nor $f'$  depend on any $P_I$, their Dirac bracket reduces to the Poisson bracket, and then there is also a weak Poisson algebra homomorphism.
\\

Next define 
\begin{equation}
H_{I}(\tau ) = H_{I}(Q^{a}(\tau ), P_{a}(\tau ), \tau ) \doteq \mathscr{O} [h_{I} ] (\tau ) \approx h_I(Q^{a}(\tau ), P_{a}(\tau ), \tau ).
\end{equation}
If $f$ is any phase space function depending only on $q_{a}, p_{a}$, but not on $T^{I}, P_{I}$, one has
\begin{equation}
\frac {\partial } {\partial \tau ^{I}} \mathscr{O} [f] (\tau ) \approx \big\{ H_{I}(\tau ),  \mathscr{O} [f] (\tau ) \big\}. \label {GEvolutn}
\end{equation}
That is, $H_{I}(\tau )$ generates the $\tau$-parametrized gauge flow of $f$ on the constraint surface. So if one specializes to a parametrization invariant dynamical system whose canonical
Hamiltonian vanishes, one may refer to the $H_{I}(\tau )$ as the ($\tau$ dependent) physical Hamiltonians. In the following we will make extensive use of (\ref{GEvolutn}),
and it is important to realize that both that PDE as well as its integrability condition hold only on-shell \cite{Dittrich_RF}.  So we will henceforth limit ourselves to on-shell physics. \\

In order to quantize the system on the reduced phase space where the classical constraints are valid, the gauge invariant canonical variables $Q^{a}(\tau ), P_{a}(\tau ); T^{I}(\tau), 
P_{I} (\tau )$
are mapped to operators $\hat{Q}^{a}(\tau ), \hat{P}_{a}(\tau ); \hat{T}^{I}(\tau), \hat{P}_{I} (\tau )$ which generate the quantum algebra $\mathfrak{U}$ with the usual equal-$\tau $ 
canonical commutation relations. Given $\mathfrak{U}$, its representation (carrier) Hilbert space $\mathscr{H}$ may be generated via the GNS construction employing any positive linear 
functional (state) on $\mathfrak{U}$. Here one is reducing phase space before quantizing; that is, all the constraints are satisfied at the classical level. We assume that for all $\tau $,
the physical Hamiltonians $ H_{I}(Q^{a}(\tau ), P_{a}(\tau ), \tau )$ are represented as densely defined self-adjoint operators on $\mathscr{H}$. \\

\section{Non-Ultralocality}

We say $h_I$ is \emph{ultralocal} if it only depends on the canonical fields or their spatial gradients (of any finite order) at the point $y(I) \in \Sigma$. If all the $h_I$ are ultralocal, then $\{h_{I} , h_{J} \} \propto \delta(y(I), y(J))$ (or its spatial derivatives) and therefore one has
$\{ H_{I} (\tau ), H_{J} (\tau) \} \approx 0$ for $y(I) \ne y(J)$.  In this case,  $H_{I}(\tau)$ and $H_{J}(\tau)$ have no common $q^{a}, p_{a}, T^{K}$ for $y(I)\ne y(J)$.
Ultralocality produces significant mathematical convenience and simplification.
Moreover, all the commonly used classical gravitational
constraint algebras, such as ADM and Holst, including possible scalar matter fields,  possess ultralocal $h_I$, see \cite{Han}. 
In fact, all the known interactions in the standard model of particle physics are 
ultralocal as well. However, a field theory is not required to be ultralocal, just that measurements so far are consistent with ultralocality. Here we keep an open mind, 
and explore the consequences of non-ultralocal $h_{I}$.   One should realize from the outset that the \emph{constraint} $\tilde{C}_{I}$ algebra remains Abelian for non-ultralocal $h_I$, even 
though neither $h_I$ nor $H_I$ will possess weakly Abelian algebras. We will see later on that the quantity $[\hat{H}_{I} (\tau ), \hat{H}_{J} (\tau )]$ for $y(I) \ne y(J)$ plays a crucial role in 
Lieb-Robinson bounds.  \\

It is sensible to first confirm that the quantum dynamics remains anomaly-free even for non-Abelian $\hat{H}_{I} (\tau )$. This means that all the classical gauge symmetries are 
faithfully reproduced 
in the quantum theory. Han \cite{Han} has proposed the condition that the $H_I$ form an Abelian algebra to be used as a definition for freedom from anomalies.  
It is demonstrated below that this is overly 
restrictive, and we
provide an alternative condition for the quantum dynamics to be anomaly-free. Following \cite{Han}, one seeks to solve the Schr\"odinger equation 
\begin{equation}
\frac{\partial}{\partial\tau _{I} } \,\hat U (\tau, \tau ') = \hat{H}_{I}(\tau)\, \hat{U}(\tau, \tau ') \label{UEvolutn}
\end{equation}
for a unitary propagator $\hat{U}(\tau, \tau ')$ between two Schr\"odinger states $\Psi(\tau), \Psi '(\tau ')$ at initial ``multi-fingered time" $\tau '$ and final value $\tau$. 
Let $\mathscr{T}$ denote the space for $\tau$, and let $c:\mathbb{R} \rightarrow \mathscr{T}$ be a path from $\tau '$ to $\tau$ parametrized by a real-valued ``external time"
$t$. We will show that $U(\tau, \tau')$ is independent of the choice of path $c$ as long as $dc^{I}/dt \ge 0$, $\forall I, t$. This is a mathematical representation of the general covariance (gauge
invariance) of the quantum dynamics. We will need a little more terminology. The real-valued fields $\tau^{I}(t)$ specify one $\tau$ at any given $t$ for each $y\in\Sigma$ and each gauge $i=0
\dots N_{c}-1$. We will refer to a single set of configurations (all $y,i$) $\{\tau^{I}(t) \}_{t}$ \emph{for all} $t$ as a ``slicing." One configuration (all $y,i$) at  \emph{one} given t is called a ``slice"
from a slicing. A gauge transformation is then a change of \emph{slicing} from $\{\tau (t)\}_{t}$ to $\{\tilde{\tau} (t)\}_{t}$; i.e., a change of multi-fingered time.
Independent from slicing invariance, $t$ can be  smoothly re-parametrized
to $t' = f(t)$ (a 1-diff). \\

Start by rewriting the $\tau$-evolution PDE (\ref{UEvolutn}) as an integral equation:
\begin{equation}
 \hat U (\tau, \tau ') = \hat{1} + \sum _{I} \int _{\tau ' _{I}} ^{\tau _{I}} \mathrm{dr}\, \hat{H}_{I}(\sigma )\, \hat{U}(\sigma ,\tau '). \label{IntUEvolutn}
 \end{equation}
 Here $\sigma $ is a variable like $\tau$ or $\tau '$ taking a value in $\mathscr{T}$ whose components are:
  \begin{equation*}
 \sigma _{J} =  
 \begin{cases}
 r , \text {for } J=I, \\
 \tau ' _{J} \text{ otherwise},
 \end{cases}
 \end{equation*}
 where $I$ is the summation index in (\ref{IntUEvolutn}), and $r$ is the (real) variable of integration.
 Iterating this leads to a Dyson expansion:
 \begin{align}
  \hat U (\tau, \tau ')   =  & \hat{1} + (-i) \sum _{I} \int _{\tau ' _{I}} ^{\tau _{I}} \mathrm{d} \sigma ^{(1)}_{I} \hat{H} _{I} (\sigma ^{(1)}) + (-i)^{2} \sum _{I,J} \int 
  _{\tau ' _{I}}^{\tau _{I}} \mathrm{d}\sigma ^{(1)}_{I}  \int _{\tau '_{J}}^{\sigma ^{(1)}_{J} }  \mathrm{d} \sigma ^{(2)}_{J} \times \nonumber \\
   & \times \hat {H} _{I}(\sigma ^{(1)})\, \hat{H} _{J} (\sigma ^{(2)}) \, \hat U (\sigma ^{(2)}, \tau ') \label{DysonL1} \\
    & = \hat{1} + \sum _{n=1} ^{\infty} (-i)^{n} \sum _{I_{1}, \dots , I_{n}} \int _{\tau ' _{I_1}} ^{\tau_{I_1}} \mathrm{d} \sigma ^{(1)}_{I_1} \int _{\tau ' _{I_2}}^{\sigma ^{(1)}_{I_2} }
    \mathrm{d}\sigma ^{(2)}_{I_2} \cdots \int _{\tau ' _{I_n}} ^{\sigma ^{(n-1)}_{I_n} } \mathrm{d}\sigma ^{(n)}_{I_n}\times \nonumber \\
    & \times \big\{ \hat{H}_{I_1} (\sigma ^{(1)})\, \hat{H}_{I_2} (\sigma ^{(2)}) \cdots 
    \hat{H}_{I_n} (\sigma ^{(n)}) \big\}, \label{DysonL2} 
    \end{align}
where $n$ is the depth of iteration, and $I_{1}, \dots , I_{n} \in \mathscr{I}$. One may concretely picture each fixed $I$ as an ``$I$-channel" from $\tau '$ to $\tau$.  
Each $\sigma^{(m)}$ is a $\tau$-vector (a point in $\mathscr{T}$), having real components $\sigma^{(m)}_{K}, K\in\mathscr{I}$. 
The $I$-subscripts on the $\hat{H}_I(\sigma )$ correspond to the $\tau _{I}$ integrated over when its corresponding argument  $\sigma_I$ is integrated.\\
    
Suppose the $\sigma ^{(n)}$ satisfy $\sigma ^{(m-1)}_{K} \ge \sigma ^{(m)}_{K}$ for all $m$ and $K$, where here $K \in\mathscr{I}$ plays a role like the index $I$ in $\tau_{I}$. We refer to this as the
\emph{ordering hypothesis} for the path through $\mathscr{T}$-space from the initial $\tau '$ to the final $\tau$.
Now consider the $n$-th term in the sum on the RHS of (\ref{DysonL2}). Suppose among the $n$ channels $I_{1}, 
    \dots , I_{n} \in \mathscr{I}$, $J_1\in\mathscr{I}$ occurs $p_1$ times, 
    $\dots , J_{M} \in \mathscr{I}$ occurs $p_M$ times, $1\le M\le n$, with $0\le p_{1}, \dots , p_{M} \le n$, and $p_{1} + \dots + p_{M} =n$. 
Under the ordering hypothesis one can recast equation (\ref{DysonL2}) as
 
 \begin{align}
 \hat U (\tau, \tau ')   =  & \hat{1} +  \sum _{n=1} ^{\infty} (-i)^{n} \sum _{1\le M\le n} \sum ^{M}_{\substack{\{p_{1} \dots p_{M}\} = 1 \\ p_{1}+\cdots +p_{M} =n }} 
 \Bigg[\prod_{k=1}^{M} \Bigg( \sum _{i(J_{k})=0}^{N_{c}-1} \int _{\Sigma} {\mathrm{d}}^{3} y(J_{k})\, \mu \big(y(J_{k})\big) \Bigg) \Bigg] _{NCC}\times \nonumber \\
 & \times \Bigg[ \prod _{k=1} ^{M}
\Bigg(  \frac {1}{p_{k} !} \int _{\tau '_{J_k}} ^{\tau _{J_k}} \mathrm{d} \tau^{(1)}_{J_k} \cdots  \int _{\tau '_{J_k}} ^{\tau _{J_k}} \mathrm{d} \tau^{(p_k)}_{J_k}\Bigg) \Bigg] \times \nonumber \\
& \times T_{\tau} \big\{\hat{H}_{I_1}(\sigma^{(1)}) \cdots \ \hat{H}_{I_n}(\sigma^{(n)})\big\}. \label{DysonF} 
\end{align}
Here the channels $I_{j} \in \mathscr{I}, j=1, \,\cdots , n$ are drawn from the $(p_{1} \times J_{1} + p_{2} \times J_{2} +\, \cdots )$ $J$'s.
The sum over the $I_1, \ldots, I_n$ has been decomposed into sums over the gauge index $i(J_k)$ and integrals over the 3-manifold $\Sigma$ with respect to the volume element
$\mathrm{d}^{3}y\, \mu(y)$, where $y$ coordinatizes $\Sigma$. The existence of a such a volume element is assured once $\Sigma$ is taken to be an oriented manifold, and it is 
not generally necessary to have a metric tensor on $\Sigma$ for that.  However, $\hat U(\tau , \tau ')$ on the LHS of  (\ref{DysonF})  
must be invariant under smooth coordinate reparametrizations of $\Sigma$, 
and this requires the volume element to be similarly invariant for scalar $\hat{H}_I$. It is straightforward to see that this can be done if it is possible to induce a metric tensor 
$g_{ab}(y)$ on $\Sigma$, 
where the volume element then takes the standard form $\mathrm{d} ^{3}y\, |\mathrm{det} \,g_{ab}(y) |^{1/2}$. $g_{ab}(y)$ does not have to be any physical metric.  
In fact, if the smooth 3-manifold $\Sigma$ smoothly 
embeds into any Euclidean space $E=\mathbb{R}^n$, then the Euclidean metric  tensor on E induces a suitable metric tensor on $\Sigma$. 
The (strong) Whitney embedding theorem states that if $\Sigma$
is $m$-dimensional Hausdorff and second countable, then $\Sigma$ smoothly embeds into $E$ for $n=2m$. Taking $\Sigma$ to have those properties, 
one then has a volume element on it to render $\hat U(\tau , \tau ')$ coordinate reparametrization invariant on $\Sigma$, moreover this is true for any such invariant volume element. 
Alternatively, taking $\Sigma$ to be a paracompact differentiable 3-manifold assures it has a Riemannian structure, whose metric tensor can then be used to construct an invariant 
3-volume element in the standard way.\\

The notation NCC on the RHS of (\ref{DysonF}) stands for ``non-collisional channels," and arises from the following considerations: Each $I$-channel has a 
gauge index $i$ and position $y$ on $\Sigma$. $i$ and $y$ are independent degrees of freedom for each channel; however, once the number of distinct channels has been fixed to be $M$, 
two channels with
the same $i$ values cannot occupy the same position $y$. That is the channels cannot ``collide," as the associated merger or splitting of channels would alter the previously fixed number of
 channels $M$, so "NCC"  can also be thought of as necessary to avoid double or under counting. The $\{ p_k\}$ give the number of $\tau$-steps or integrations along each distinct channel $J_k$. 
Any single $\sigma ^{(k)} \in \mathscr{T}$ has the components: \emph{one} from the sequence $(\tau^{(1)}_{J_1}, \,\cdots , \tau^{(p_1)}_{J_1})$ of real integration variables in channel $J_1$;  $\,\dots\,$; 
\emph{and}  any \emph{one} from the sequence $(\tau^{(1)}_{J_M}$, $\cdots$ , $\tau^{(p_M)}_{J_M})$ of integration variables in channel $J_M$; \emph{and} 
(if not integrated) that component of $\sigma^{(k)}$ is set equal to the corresponding component of the initial $\tau '$.
Because of the ordering hypothesis, $\sigma ^{(k-1)}$ differs from $\sigma ^{(k)}$ at only one index value $J$, where $\sigma ^{(k-1)}_{J}\ne \sigma ^{(k)}_{J}$, and then $I_{k}=J$.
The ``$\tau$-ordering operator" $T_{\tau}$ acts to order the $\tau$ arguments in each $I$-channel independently so that within each channel $J$ they increase from right to left:
$\sigma^{(n)}_{J}\le \cdots \le \sigma^{(2)}_{J} \le \sigma^{(1)}_{J}, \forall J$.
Notice that the ordering hypothesis has allowed us to remove the path-dependent limits on the multiple $\tau$-integrations.\\

Equation (\ref{DysonF}) 
sums (averages) over all the paths in $\mathscr{T}$ from $\tau '$ to $\tau$ by 
advancing monotonically in $\tau_I$ within each $I$-channel independently, stepwise over the $n$ integrations, from initial $\tau '$ to final $\tau$, as enacted by the $T_{\tau}$ operation.
The averaging over the paths from the initial to final $\tau$-slices is carried out for each fixed $n, M$, first at fixed (NCC) channel configuration $\{ (i(J_k), y(J_k)) \}_k$ 
and fixed stepping configuration
$\{ p_k\}_k $ by the corresponding $\tau$-integrations. Then the channel configuration and stepping configuration are separately averaged at fixed $n, M$ 
(the result is independent of which of the latter two averages is performed first).
Now suppose we have some ``external time" parameter $t$ so that $\tau _{I}=\tau_{I} (t)$. 
As long as the path  $c^{I}(t) \doteq \tau_{I}(t)$ between the fixed initial and final $\tau$ obeys $\mathrm{d} c^{I}(t) /\mathrm{d}t \ge 0 $ for all $I\in\mathscr{I}$ and
relevant values of $t$, one fulfills the ordering hypothesis. This mild monotonicity condition is consistent with the physical picture of multi-fingered time 
as ``flowing forwards everywhere."  The overall sign of the monotonicity condition may be reversed, so mutli-fingered time then globally flows "backwards," however the choice of that
sign does not affect the conclusions. \\

By adding over all such ways of channel-wise monotonic advancement from $\tau '$ to $\tau$, the overall RHS of (\ref{DysonF}) is insensitive to a change of slicing (gauge)
 $\tau_{I}(t) \rightarrow \tilde{\tau}_{I}(t) $ for fixed initial $\tau '$ and final $\tau$, and $ \hat U (\tau, \tau ') $ depends only on those initial and final configurations.
This happens regardless of the commuting or non-commuting 
properties of the $\hat{H}_{I}(\tau )$. It occurs because the monotonicity condition takes care of the required $\tau _{I}$-ordering within each $I$-channel separately ($T_{\tau}$
becomes a $t$-ordering), 
and the operator orderings among different $I$-channels (inside $T$) on the RHS are averaged over (as a sum over monotonic paths or slicings between the fixed initial $\tau '$
and final $\tau$ 
configurations). This absence of path or slicing dependence of  $\hat{U}(\tau, \tau ')$ is the relational framework manifestation of general covariance (gauge invariance) for the quantum dynamics: we have obtained 
freedom from anomalies for $\tau$-paths monotonic in an external time parameter. Under these conditions the propagator $\hat{U}$ more resembles the familiar one from standard 
 (fixed background geometry) quantum field theory.  In the 
absence of monotonicity, the use of the components of $\tau$ as integration variables, such as in equations (\ref{IntUEvolutn}),  (\ref{DysonL1}), and (\ref{DysonL2}) becomes ill-defined: 
Some $I$-channels could then have ranges of $\tau_I$ which are traversed multiple times in both senses as $t$ advances.
This accords with the intuition that these locally negative lapses among generally positive ones are indeed somehow physically anomalous. 
 In fact, a similar notion in a different guise was used in \cite{Oriti_Livine} to generate ``causal"  spin-foam vertices and amplitudes. 
 Monotonicity may be viewed as the relational framework analog of global hyperbolicity on Lorentzian manifolds;
however unlike the latter, monotonicity assumes no background causal structure.\\

If one specializes to the ultralocal case, so $\hat{H}_{I}(\tau)$ and $\hat{H}_{J}(\tau)$ have no common $\hat{Q}^{a}, \hat{P}_{a}, \hat{T}^{K}$ for $y(I)\ne y(J)$, and to the gauge diagonal case, 
for which the same thing occurs when $i(I)\ne i(J)$, then $[\hat{H}_{I}(\tau), \hat{H}_{J}(\tau)] \approx 0$ for any $I\ne J$. Consequently the $T_{\tau}$-ordered product in (\ref{DysonF}) factors into 
separate $T_{\tau _I}$ for each $I$-channel, and $\hat{U} = \prod _{I} \hat{U}_{I}$, as obtained earlier by Han \cite{Han}.  \\

It is also interesting to compare the $T_{\tau}$-ordering in (\ref{DysonF}) with the well-known $T$-ordering from standard quantum field theory. Weinberg \cite{Weinberg_1} gives the following 
Dyson expansion for the S-matrix (operator) in Minkowskii spacetime or special relativity:
\begin{equation}
\hat{S} = \hat{1} + \sum _{n=1}^{\infty} \int \mathrm{d} ^{4} x_{1} \cdots \mathrm{d} ^{4} x_{n} \, T\{ \hat{\mathscr{H}}(x_{1}) \cdots \hat{\mathscr{H}}(x_{n})\}, \label{SMatrix}
\end{equation}
where $\hat{H} =\hat{H}_{0} + \hat{V}$, $\hat{V}(t) = \exp (i\hat{H}_{0}t)\, \hat{V}  \exp (-i\hat{H}_{0}t)$, $\hat{V}(t) =\int \mathrm{d}^{3}x \,\hat{\mathscr{H}}(\vec{x}, t)$ in the interaction picture, 
with $\hat{H}_{0}$ the free-Hamiltonian and $\hat{V}$ the interaction. This is globally Lorentz invariant except for the $T$-ordered operator product. The $T$-order 
of two spacetime points
 $ x_{1}$ and $x_{2}$ (the order of their 0-coordinates) 
 is globally Lorentz invariant unless $(x_{1}-x_{2})^{2} > 0$ ($x_{1}-x_{2}$ space-like), so (\ref{SMatrix}) introduces no special frame if (but not only if) the 
 $\hat{\mathscr{H}}(x)$ commute at space-like distances. While this is often referred to as a kind of causality condition, in this sense it arises from the invariance of $\hat{S}$ that occurs 
 because global Lorentz transforms $x \rightarrow x' = \Lambda x $ alter the spatial and temporal components of 4-vectors and so can then re-order the $T$ sequence among the $\{x_{n}\}$.
 By contrast, in the relational framework representation just discussed, $t$ and all the $\tau _{I}$ increase smoothly within both slicings $\tau _{I}(t)$ and $\tilde{\tau}_{I}(t)$ connecting the fixed initial and final
 $\tau$'s, and the gauge transform is no longer simply related to the spatial coordinates $y$. Thus the relational framework has no built-in analog of special relativity's ``causality from global Lorentz invariance." 
 The $T_{\tau}$
 in equation (\ref{DysonF}) will not generally lead to ``causality" by itself, regardless of whether the constraints are ultralocal or not. \\
 
 The quantity $[\hat{H}_{I}(\tau ), \hat{H}_{I}(\tau )]$ for $y(I) \ne y(J)$  is important in the  subsequent sections. Therefore we spend some time to study it
 as well as its classical precursor $\{ H_{I}(\tau ), H_{I}(\tau )\}$. \\
 
 We start by computing classically,
 \begin{equation}
\big\{H_{I}(\tau ), H_{J}(\tau ) \big\} \approx \mathscr{O}\big [\{ h_{I},h_{J} \}_{D} \big](\tau ) = \mathscr{O} \big[\{ h_{I},h_{J} \} \big] (\tau ), \label{HHPB_1}
\end{equation}
where the first relation has used equation (\ref{PBHomomorph}), and the second equality used that the $h_{I}$ do not depend on any $P_{J}$ inside some phase space neighborhood of the constraint
surface. We also have $\tilde{C}_{I} = P_{I} + h_{I}$ with $\{\tilde{C}_{I}, \tilde{C}_{J}\} = 0$ (Recall the $\tilde{C}_{I}$ form a strongly Abelian constraint algebra), and one derives
\begin{equation}
0=\{P_{I} + h_{I} , P_{J} + h_{J} \} = \{ h_{I} , h_{J}\} + \{P_{I}, h_{J}\} +\{h_{I}, P_{J}\} = \{ h_{I} , h_{J}\} + \frac {\delta h_{J} } {\delta T^{I}} - \frac {\delta h_{I} } {\delta T^{J}}.
\end{equation}
Thus,
\begin{equation}
\big\{ H_{I}(\tau ), H_{J}(\tau )\big\} \approx \mathscr{O} \Bigg[\frac {\delta h_{I} } {\delta T^{J}} - \frac {\delta h_{J} } {\delta T^{I}}\Bigg] (\tau ). \label{HPBrac2}
\end{equation}
We use the same sign conventions as Han \cite{Han}, namely
\begin{align}
\{ p_{a} , q^{b} \} & = (+) \delta ^{b}_{a} \nonumber \\ 
\{ P_{I}, h_{J} \} & = (+) X_{P_I} \cdot h_{J} = \frac {\delta h_{J}} {\delta T^{I}}  \label{HanConvs} \\
X_{f} & = \frac {\delta f} {\delta p_{i}} \frac {\delta}{\delta q^{i}} - \frac {\delta f} {\delta q^{i}} \frac {\delta}{\delta p_{i}} .\nonumber 
\end{align}
One also has from (\ref{HHPB_1}):
\begin{align}
\mathscr{O} \big[\{ h_{I}, h_{J} \} \big](\tau ) & = \mathscr{O} \big[\{\tilde{C}_{I} , h_{J}\}\big](\tau ) - \mathscr{O} \big[\{P_{I} , h_{J}\} \big](\tau ) \nonumber \\
& = \sum _{\{k_{I}\}} \Bigg( \prod _{J} \frac {(\tau ^{J} - T^{J})^{k_J}} {k_{J} !} \Bigg) \Bigg( \prod _{J} (X_{J})^{k_{J} } \Bigg) \tilde{X}_{I} \cdot h_{J}
     - \mathscr{O} \Bigg[\frac{\delta h_{J}}{\delta T^{I} } \Bigg] (\tau ) \nonumber \\
 & \approx \sum _{\{k_{I}\}} \Bigg( \prod _{J} \frac {(\tau ^{J} - T^{J})^{k_J}} {k_{J} !} \Bigg) \tilde{X}_{I} \Bigg( \prod _{J} (X_{J})^{k_{J} }\Bigg) \cdot h_{J}
     - \mathscr{O} \Bigg[\frac{\delta h_{J}}{\delta T^{I}} \Bigg] (\tau )  \nonumber   \\
 & \approx \sum _{\{k_{I}\}} \Bigg( \prod _{J} \frac {(\tau ^{J} - T^{J})^{k_J}} {k_{J} !} \Bigg) {X}_{I} \Bigg( \prod _{J} (X_{J})^{k_{J} } \Bigg) \cdot h_{J}
     - \mathscr{O} \Bigg[\frac{\delta h_{J}}{\delta T^{I}} \Bigg] (\tau )     \nonumber \\
 & = \frac {\partial H_{J}(\tau )} {\partial \tau^{I} }   - \mathscr{O} \Bigg[\frac{\delta h_{J}}{\delta T^{I} }\Bigg] (\tau ).  
 \end{align} 
Consequently,
 \begin{equation} 
 \big\{H_{I}(\tau ), H_{J}(\tau )\big\} \approx \frac {\partial H_{J}(\tau )} {\partial \tau^{I} }   - \mathscr{O} \Bigg[ \frac{\delta h_{J}} {\delta T^{I} } \Bigg] (\tau ) \label{HPBrac3}. 
 \end{equation}
 Combining (\ref{HPBrac2}) and (\ref{HPBrac3})  one arrives at
 \begin{equation}
 \frac {\partial H_{J}(\tau )} {\partial\tau^{I}} \approx \mathscr{O} \Bigg[ \frac{\delta h_{I} }{\delta T^{J} } \Bigg] (\tau ). \label{dHdTau} 
 \end{equation}
 
Which canonical variables enter $\mathscr{Q}_{IJ} \doteq \{H_{I}(\tau ), H_{J}(\tau )\}$? Since $H_{I}(\tau) =\mathscr{O}[h_{I}](\tau )$ and near shell $h_{I}$ is independent of $P_{K}$'s,
so $H_{I}$ has no $P_{K}$'s and neither does $\mathscr{Q}_{IJ}$. How about the $(q^{a}, p_{a})$ variables? Let $f$ be an arbitrary phase space function solely dependent on the
 $(q^{a}, p_{a})$'s (and not containing any $T^{K}, P_{K}$ variables). One then has,
  \begin{align}
\{H_{I}(\tau ), \mathscr{O}[f](\tau )\} & \approx \mathscr{O} [ \{ h_{I}, f\} _{D}](\tau ) =  \mathscr{O} [ \{ h_{I}, f\} ](\tau )\nonumber \\
&  \approx \mathscr{O}[ \{\tilde{C}_{I}, f\} ] (\tau ) \approx \mathscr{O} [X_{I} \cdot f] (\tau ). \label{Iter}
\end{align}
In the second equality we have used that $h_{I}, f$ are independent of $P_{K}$ variables, in the third that $f$ contains no $T^{I}$ variables, and in the fourth that $\tilde{C}_{I} =
C'_{I}+\mathcal{O}(C^{2})$ and $\{f_{1}, f_{2}\} = X_{f_{1} } \cdot f_{2}$. One would like to  iterate equation (\ref{Iter}) to yield
\begin{align}
\{ H_{I}(\tau ), \{ H_{J}(\tau ),\mathscr{O}[f] (\tau ) \}\} \,& {} ^{\, ?}_{\approx} \, \{ H_{I}(\tau ), \mathscr{O}[X_{J} \cdot f](\tau )\} \nonumber \\
& \approx \mathscr{O}[(X_{I} \cdot X_{J}) \cdot f] (\tau )  \approx
\mathscr{O} [(X_{J} \cdot X_{I}) \cdot f](\tau ), \label{HHOfBrac5}
\end{align}
where in the last equality the weak commutativity of the $X_{I}$ has been used.
If this were true, then by using the Jacobi identity and that $f$ is an arbitrary phase space function of the $q^{a}, p_{a}$ variables, one would ``find" that 
$\{ H_{I}(\tau ), H_{J}(\tau )\} $ could not be a phase space function of any $q^{a}, p_{a}$. However, the step marked with ?  is invalid because there one is using the weak equation 
(\ref{Iter}) \emph{inside} a Poisson bracket.  
Such a maneuver is \emph{inadmissible}, as no weak equation may be used before evaluating Poisson brackets. Hence $\{ H_{I}(\tau ), H_{J}(\tau )\} $ may still be a phase 
space function of $q^{a}, p_{a}$ as well as the clock variables. \\
  
  Suppose the constraints are ultralocal, meaning $h_{I}$ only contains $q^{a}, p_{a}$ with $y(a)=y(I)$ and clocks $T_{K}$ with $y(K)=y(I)$. Then $\delta h_{J}/\delta T^{I} \propto \delta
  (y(I), y(J))$, so by (\ref{HPBrac2}) $\{ H_{I}(\tau ), H_{J}(\tau )\} \approx 0$ for $y(I)\ne y(J)$. Hence to obtain $\{ H_{I}(\tau ), H_{J}(\tau )\} \ne 0$ for $y(I)\ne y(J)$
  requires non-ultralocal constraints. \\
  
  What about spatial parity $(P)$ and time reversal $(T)$ symmetries? Let $\mathscr{Q}^{jk}_{xy}\,$$ \doteq $$\{ H_{I}(\tau ), H_{J}(\tau )\} $ where $j=i(I)$, $k=i(J)$, $x=y(I)$, and $y=y(J)$. One has
  $\mathscr{Q}^{jk}_{xy} =- \mathscr{Q}^{kj}_{yx}$ by anti-symmetry of the Poisson bracket. One might be concerned that when $x\ne y$, for $\mathscr{Q}^{jk}_{xy}$ to be non-vanishing could 
  require violation of $P$-symmetry (in addition to non-ultralocality); i.e., that  $\mathscr{Q}^{jk}_{xy} $ could acquire a non-zero spatially odd piece. Such a violation
   is only necessary provided $\mathscr{Q}^{jk}_{xy} =
\mathscr{Q}^{kj}_{xy}$, that is the constraints are ``gauge symmetric," meaning they satisfy
\begin{equation}  
 \frac{\delta h^{k}_{x} }{\delta T^{j}_{y}} = \frac{\delta h^{j}_{x} }{\delta T^{k}_{y}},     \;\;\; (\text{Gauge Symmetry}) \label{GaugeSymm}
 \end{equation}
which does not generally hold. So a violation of $P$-symmetry is not generally necessary to obtain $\{ H_{I}(\tau ), H_{J}(\tau )\} \ne 0$ for $y(I)\ne y(J)$ if there is non-ultralocality.
An examination of the formal
power series for $\mathscr{O}[f](\tau)$ (\ref{SeriesExpsn}) similarly shows that ones does not require a violation of $T$-reversal symmetry either. The absence of requiring
 $P$- and/or $T$-violation to obtain
$\{ H_{I}(\tau ), H_{J}(\tau )\} \ne 0$ for $y(I)\ne y(J)$ is reassuring since the gravitational interaction is not expected to violate those symmetries. \\

We now present a very simple toy example of non-ultralocality. Suppose the Hamiltonians have the special form
\begin{equation}
h^{i}(x) = \bar{h} ^{i}(x) + \int _{\Sigma} \mathrm{d}^{3}y\; K^{ij}(q^{a}(x), p_{a}(x), q^{b}(y), q^{b}(y); x,y)\; T^{j}(y), \label{LTA}
\end{equation}
with the non-ultralocal term chosen linear in the clock variables for simplicity. We take the clock variables as 3-diff scalars and the kernel $K^{ij}$ to be a weight-one 3-density. The first term on the 
RHS is ultralocal and does not contribute to $\{ H_{I}(\tau ), H_{J}(\tau )\}$.  One finds
\begin{align}
\mathscr{Q}(\tau )& \doteq \{ H^{x}_{l} (\tau ), H^{z}_m(\tau )\} \approx \mathscr{O}\bigg[\frac {\delta h^{x}_{l}} {\delta T^{z}_{m}} - \frac {\delta h^{z}_{m}} {\delta T^{x}_{l}}\bigg](\tau ) \nonumber  \\
& \approx \mathscr{O} \big[K^{lm}(q^{a}(x),p_{a}(x), q^{b}(z), p_{b}(z); x,z) - K^{ml}(q^{b}(z),p_{b}(z), q^{a}(x), p_{a}(x); z,x)\big] (\tau ). 
\end{align}

Next we examine the $\tau$ and external time gauge flow in more detail. This is the bridge we will need to cross to get to the Lieb-Robinson bounds.
From the classical gauge flow equation (\ref{GEvolutn}) and again following Han's sign conventions that $\{P_a, Q^{b}\} \approx \delta ^{b}_{a}$, $[\hat{P}_{a}(\tau ), \hat{Q}^{a}(\tau )] =
(-i) \{P_{a}(\tau ), Q^{b}(\tau ) \} = (-i) \delta ^{b}_{a}$ ($\hbar$ is set to unity), one has the on-shell quantum gauge flow equation:
\begin{equation}
\frac {\partial}{\partial \tau _{I} } \hat{\mathscr{O}} [f](\tau ) \approx (i) \Big[\hat{H}_{I}(\tau ),  \hat{\mathscr{O}} [f](\tau )\Big], \label{QntmTauEvolutn} 
\end{equation}
where $f$ may depend on $q^{a}, p_{a}$ canonical variables but not the $T^{I},P_{I}$ types. Adopting the Ansatz 
\begin{equation}
\hat{\mathscr{O}} [f](\tau ) =\exp \big[i\hat{M}(\tau )\big] \big(\hat{\mathscr{O}} [f](0)\big) \exp \big[-i\hat{M}(\tau )\big]  \label{Ansatz} 
\end{equation}
for some self-adjoint operator $\hat{M}$ independent of $f$, one infers from the gauge flow equation that
\begin{equation}
\Bigg[\frac{\partial\hat{M}(\tau )} {\partial \tau^{I} } - \hat{H}_{I}(\tau ), \hat{\mathscr{O}} [f](\tau ) \Bigg] \approx 0.
\end{equation}
Since $f$ is an arbitrary phase space function of $q^{a}, p_{a}$, one has that on-shell ${\partial\hat{M}(\tau )} /{\partial \tau^{I} } - \hat{H}_{I}(\tau )$ may depend
on the $\hat{T}^{K}$ but not on the $\hat{q}^{a}, \hat{p}_{a}$ or $\hat{P}_{K}$. Here we will assume the simplest case, that is
\begin{equation}
\frac{\partial\hat{M}(\tau )} {\partial \tau^{I} }  \approx \hat{H}_{I}(\tau ), \label{MTauEqn}
\end{equation}
which is sufficient, but not necessary. Now decompose $\hat{M}$ as
\begin{equation}
\hat{M}(\tau ) = \sum _{Z} \hat{M}_{Z}(\tau ),
\end{equation}
where $Z$ is called a ``patch" and is just the support of $\hat{M}_{Z}$ on $\Sigma$. The motivation behind this is as follows: Each $I\in\mathscr{I}$ contains a continuous
spatial coordinate $y(I)$ as well as
a discrete gauge index $i(I)$. For nonlocal ${h}_{I}$ and ${H}_{I}$, besides the canonical variables at $y(I)$ (called the ``central site") there are other ``nearby" canonical variables
living at $y' \ne y(I)$ which also enter $h_{I}$ and $H_{I}$ as their ``entourage." We define the patch $Z(I)$ to consist of the central site $y(I)$ together with all those nearby $y'$ where its 
entourage reside. We assume the patches to be bounded and not to take up all of $\Sigma$. Denoting the central site of $Z$ by $y_{c}(Z)$ we can set
\begin{equation}
H_{Z} \doteq H_{Z(I)} = H_{I} |_{y(I) = y_{c}(Z) }.
\end{equation}
Using (\ref{MTauEqn}) with fixed $I$,
\begin{equation}
H_{Z(I)}(\tau ) \approx \sum _{X} \partial M_{X}(\tau )/\partial \tau ^{I}. 
\end{equation}
Later on we will be more interested in the external time $t$ gauge flow of $\hat{\mathscr{O}} [f](\tau )$ than in the $\tau$-flow, so we let
\begin{align}
\mathbb{\hat{H}}(t)  & \doteq \sum_{I} \frac{\partial \tau ^{I}(t)} {\partial t} \hat{H}_{I}(\tau ) \approx \sum _{I}  \frac{\partial \tau ^{I}(t)} {\partial t} \sum _{X} \frac {\partial \hat{M}_{X}(\tau )}
{\partial \tau ^{I}} = \frac {\partial\hat{M}(t)} {\partial t} \label {HBBDefn}\\
\hat{M} (t) &  \doteq \sum _{X} \hat{M}_{X}(\tau (t)).
\end{align}
It then follows from the $\tau$ gauge flow equation (\ref{QntmTauEvolutn}) that
\begin{align}
\frac{d}{dt} \hat{\mathscr{O}} [f] (t) = \sum _{I} \frac{\partial \tau ^{I}(t)} {\partial t} \frac {\hat{\mathscr{O}}[f](\tau(t)) } {\partial\tau^{I}} & = (i) \Bigg[\sum _{I}  \frac{\partial \tau ^{I}(t)} {\partial t} \hat{H}_{I}(\tau ) , \hat{\mathscr{O}} [f] (\tau) \Bigg]   \nonumber \\
& = (i) \Big[ \mathbb{ \hat{H}}(t), \hat{\mathscr{O}} [f] (\tau (t)) \Big].  \label{BBHaction} 
\end{align}
As expected, one sees that $\mathbb{\hat{H}}(t)$ generates gauge flow in external time.
More explicitly, setting $\hat{\mathscr{O}} [f] (t) \doteq \hat{\mathscr{O}} [f] (\tau (t))$, (\ref{BBHaction}) implies
\begin{equation} 
\lim _{\epsilon\rightarrow 0} \hat{\mathscr{O}}[f](t+\epsilon) = \hat{\mathscr{O}}[f](t) + i\epsilon\, \bigg[\hat{\mathbb{H}}(t), \hat{\mathscr{O}}[f](t) \bigg] + \mathcal{O}(\epsilon ^{2}) = 
\exp \big(i \epsilon\,\hat{\mathbb{H}}(t) \big) \hat{\mathscr{O}}[f](t) \exp \big(-i\epsilon\,\hat{\mathbb{H}}(t)\big).
\end{equation}
We would like to put $\hat{\mathbb{H}}(t) \doteq \sum _{Z} \hat{H}_{Z} (t)$ and figure out what $ \hat{H}_{Z} (t)$ is, hence by (\ref{HBBDefn}) 
\begin{equation}
\sum _{Z} \hat{H}_{Z}(t) \doteq \hat{\mathbb{H}}  =  \sum_{I} \frac{\partial \tau ^{I}(t)} {\partial t} \hat{H}_{I}(\tau ) = \sum _{Z} \sum _{I: Z(I)=Z}  \frac{\partial \tau ^{I}(t)} {\partial t} \hat{H}_{I}(\tau )=
\sum _{Z} \Bigg( \frac{\partial \tau ^{I}(t)} {\partial t} \hat{H}_{I}(\tau ) \Bigg) _{Z(I)=Z},
\end{equation}
where the last equality follows from the fact that there is only one patch $Z(I)$ with central site $y(I)$ corresponding to $Z$: $y(I) = y_{c}(Z)$. Consequently,
\begin{equation} 
\sum _{Z} \hat{H}_{Z}(t) = \sum _{Z} \sum _{i=0}^{N_{c}-1} \Bigg( \frac {\partial \tau ^{i}(y_{c}(Z), t)} {\partial t} \Bigg) \hat{H}_{i}^{y_{c}(Z)}(\tau (t)).
\end{equation}
And pulling it all together:
\begin{align}
 \hat{\mathbb{H}} (t)  & \doteq \sum _{Z} \hat{H}_{Z}(t) \label{PatchyA}\\
 \hat{H}_{Z}(t) & =  \sum _{i=0}^{N_{c}-1} \Bigg( \frac {\partial \tau ^{i}(y_{c}(Z), t)} {\partial t} \Bigg) \hat{H}_{i}^{y_{c}(Z)}(\tau (t)). \label{Patchy} 
 \end{align}
 $\hat{\mathbb{H}}(t)$ is a ``patchy" Hamiltonian generating relational framework gauge flow in external time, whose patches $Z$ are based on the non-ultralocality of the original $h_{I}$. 
 This patchy representation of $t$ gauge flow unlocks the door to applying the Lieb-Robinson bounds to be introduced in section 4 below. \\
 
 Before immediately moving on to the Lieb-Robinson bound, we will need to know a little more about $[\hat{H}_{X}(t), \hat{H}_{Y}(t)]$. Here we discuss
 its classical counterpart $\mathscr{Q} \doteq \{ H_{X}(t), H_{Y}(t)\} $ to gain some intuition about  it before proceeding.  \\
 
 $\mathscr{Q}$ is built from 
 \begin{equation}
 \big\{ H^{y_{1}}_{i_{1}} , H^{y_{2}}_{i_{2}}\big\} \approx \mathscr{O} \Bigg[ \frac {\partial h^{y_{1}}_{i_{1}} } {\partial T^{y_{2}}_{i_{2}} } - \frac {\partial h^{y_{2}}_{i_{2}} } 
 {\partial T^{y_{1}}_{i_{1}} } \Bigg]  (\tau ).
 \end{equation}
Denote
 \begin{equation}
 f^{i_{1}}_{i_{2} } (x,y) \doteq \mathscr{O}\Bigg[  \frac {\partial h^{x}_{i_{1}} } {\partial T^{y}_{i_{2}} } \Bigg] (\tau).
 \end{equation}
 We have $\mathscr{Q}(x,y,t) \doteq \{H_{X}(t),H_{Y}(t)\} = -\mathscr{Q}(y,x,t)$ where $x=y_{c}(X)$ and $y=y_{c}(Y)$, and thus
 \begin{equation}
 \mathscr{Q}(x,y,t) = \sum_{i_{1},i_{2}=0}^{N_{c}-1} \Bigg[ \frac{\partial\tau_{i_{1}}(x)} {\partial t}  \frac{\partial\tau_{i_{2}}(y)} {\partial t} \Bigg] \big( f^{i_{1}}_{i_{2}} (x,y) - 
 f^{i_{2}}_{i_{1} } (y,x) \big). \label{QQQ}
 \end{equation}
 \\
 As a reminder,  $\mathscr{Q}$  may be a phase space function of the $T^{K}, q^{a}, p_{a}$ variables, but not $P_{K}$ variables, or it could just be a phase space constant, and so include 
 $\tau$'s.
As a simple example, here we will try to construct $f$ from just the $\tau$'s. We take $f^{i}_{j}(x,y)$  to be $(x,y)$ symmetric (P-conserving)
 and $(i,j)$ anti-symmetric,
 such as $f^{i}_{j}(x,y) = (\tau_{i}(x)-\tau_{j}(x)) + (\tau_{i}(y)-\tau_{j}(y))$. Then for the sum in (\ref{QQQ}) to be non-vanishing, we have to $(i_{1},i_{2})$ anti-symmetrize the brackets in that 
 expression to obtain
 \begin{equation}
 \mathscr{Q}(x,y,t) =  \sum_{i_{1},i_{2}=0}^{N_{c}-1} \Bigg[ \frac{\partial\tau_{i_{1}}(x)} {\partial t}  \frac{\partial\tau_{i_{2}}(y)} {\partial t}   - \frac{\partial\tau_{i_{2}}(x)} {\partial t}  \frac{\partial\tau_{i_{1}}(y)} {\partial t}  \Bigg]   f^{i_{1} }_{i_{2} } (x,y),
 \end{equation} 
 which is overall $(x,y)$ anti-symmetric as required. So as $y\rightarrow x$, $\mathscr{Q} \rightarrow 0$, but away from $x=y$, $\mathscr{Q}$ is non-vanishing. We expect it to decay as 
 $H_{X}$'s and $H_{Y}$'s patches $X$ and $Y$ cease to overlap, but that behavior is not well captured by this toy model for   $\mathscr{Q}$.\\
 
 Lieb-Robinson bounds were originally intended to study spin systems imbedded in a solid-state lattice, so they are naturally discretized. This lattice may be extended to include a general 
 network and is not limited
 to a periodic tessellation of 3-space by polyhedra. Discretization achieves significant mathematical simplifications, so we will follow that approach in this initial investigation of relational framework with non-ultralocal 
 constraints. We discuss the limitations
 and issues related to discretization and its continuum limit in section 7. A continuum approach will be left for future research. \\
 
 Here we describe the discretization of the 3-manifold $\Sigma$ into a (generalized) lattice $\Lambda$. Associated with each lattice site $j\in \Lambda\subset\Sigma$ is a $D$-dimensional 
 Hilbert space. Unless 
 otherwise indicated $\Lambda$ will have finite size (cardinality); alternatively, $\Lambda$ may be taken to be a finite sub-lattice of some countable lattice $\Gamma\subset\Sigma$. The Lieb-Robinson bound 
 does not depend on the
 dimensionality $D$, and the Hilbert space for for the entire system is taken to be the tensor product of the site-based spaces. Capital Latin letters from the end of the alphabet 
 (previously referring to Hamiltonian patches) will now denote sets of lattice sites,
 and $|X|$ designates the cardinality of $X$. 
 We say an operator $\hat{O}$ is supported on a set $Y$ of sites if  $\hat{O}$ may be expressed as $\hat{O} = \hat{1}_{\Lambda\setminus Y} \otimes \hat{P}$, 
 where $ \hat{1}_{\Lambda\setminus Y}$ 
 is the identity operator on sites not in $Y$, and $\hat{P}$ is an operator defined on $Y$.  In the following sections we will be most interested in the complete (Dirac) 
 observables $\mathscr{O}[f](\tau )$, where $f$ is a phase space function
 containing neither $T^{I}$ nor $P_{I}$ canonical variables, and use the unitary Hamiltonian patchy gauge flow (\ref{BBHaction}), (\ref{PatchyA}), and (\ref{Patchy}). 
 We assume we can take the discretized $\hat{H}_{I}(\tau ) \approx \hat{H}^{i}_{j}(Q^{a}(\tau), P_{a}(\tau ), 
 \tau )$ as a self-adjoint operator on the Hilbert space which is the tensor product of Hilbert spaces over lattice sites $j(a)$ included in its arguments. \\
 
 We must take a moment to carefully resolve any potential issues that might arise from discretizing operators like the $\hat{\mathscr{O}}[f](\tau)$, and to confirm that the discretized equations 
 behave as expected, especially from a gauge-flow point of view. To this end, we have to define what precisely is meant by the spatial discretization $\Delta$ of an continuum operator
 constructed as a sum of products of the 
 canonical variables. The discretization map $\Delta$ is defined to act linearly with respect to any sum of operators. Acting on a product of continuum operators, $\Delta$ annihilates (``apodizes" or cuts off) any 
 product which contains one or more factors of canonical variables that are not on the lattice. From this definition follows $\Delta(\hat{A}\hat{B}) = \Delta(\hat{A})\Delta(\hat{B})$. 
The procedure we follow is to "Diracify" first by constructing the continuum Dirac operator  $\hat{\mathscr{O}}[f](\tau)$ from $f$, and then to discretize by acting with $\Delta$. One seeks to 
 demonstrate that the continuum gauge-flow equation (\ref{QntmTauEvolutn}) holds when all operators are replaced by their discretized images under $\Delta$; i.e., that 
 $\Delta$ is a gauge-flow
 homomorphism. This is made easier after one
notes that the continuum gauge-flow equation (\ref{QntmTauEvolutn}) is an operator equation with both sides (weakly) equal to the operator corresponding to the classical expression 
\begin{equation}
\sum_{k_{J}=0}^{\infty} \Big( \prod_{J} \frac {(\tau^{J} - T^{J})^{k_{J}} } {k_{J}!} \Big)\, X_{I} \, \Big( \prod _{J} (X_{J})^{k_{J}} \Big)\, \cdot f,
\end{equation}
see \cite{Han} equation 2.13. Then by restricting the free index $I$ to have $y(I)\in \Lambda$, i.e. to be on the lattice, and applying the $\Delta$ map, it is straightforward to show that
\begin{equation}
\frac {\partial } {\partial \tau ^{I}} \Delta ( \hat{\mathscr{O}} [f] (\tau ) ) \approx (i) \Big[ \Delta\big(\hat{H}_{I}(\tau ) \big),  \Delta \big( \hat{\mathscr{O}} [f] (\tau ) \big) \Big]. \label{DiscGFlow}
 \end{equation}
 One has to interpret the $\tau^{J}=\tau^{J}\,\hat{1} $ terms as $\Delta (\hat{1}) = \hat{1}_{\Lambda} = \otimes _{j\in\Lambda} \hat{1}_{j}$, where $\hat{1}_{j}$ is the identity operator
 on the Hilbert space at site $j$. This way all the $J$'s appearing in the sums inside (\ref{DiscGFlow}) are on lattice, and there is no on-shell operator
 flow to/from the lattice from/to non-lattice-sites. It is also simple to show that $[\Delta ( \hat{C}_I), \Delta(\hat{C}_J)] \approx 0$, so the discretized constraints are weakly Abelian.
 Henceforth we drop the $\Delta$ whenever it is clear from the context that we are discussing a discretization. \\
 
 If $i,j$ are lattice sites on $\Sigma$, the Lieb-Robinson bound require a 3-metric $d(i,j)$. If $A,B$ are sets of lattice sites, for future use we define 
 \begin{align}
 d(A,B)\doteq\mathrm{dist}(A,B)  & \doteq \min_{i\in A, j\in B} \, d(i,j) \\
 \mathrm{diam}(A) &  \doteq \max_{i,j \in A} \,d(i,j). 
 \end{align}
  
 When Lieb-Robinson bounds were first applied to solid-state spin systems, introducing the static metric $d(i,j)$ was innocuous, however in applying Lieb-Robinson bounds to gravitational physics there are several issues of
 serious concern.
 Already at the purely classical level, $d$ will acquire a dependence on geometric variables included in the $Q^{a}(\tau )$, 
 so if one has a continuously varying external time parameter $t$, $d(i,j)$ will inherit a 
 continuous $t$ dependence as well,
 while $d$ still describes a discretized 3-geometry of $\Sigma$. We will show in Section 5 how this $t$ dependence can be accommodated within the Lieb-Robinson bound. Still at the classical level, in the continuum $d(x,y)$ for $x,y\in\Sigma
 $ could be taken as the
 proper geodesic distance between $x$ and $y$. But once $\Sigma$ has been discretized, the voxelated classical 3-geometric information and the replacement of the PDE describing a 
 geodesic by a finite difference
 equation will introduce a classical discretization ``error"  into $d(i,j)$. Of course, one expects  this classical error to become negligible in the limit where the (proper)
 lattice cell size becomes much smaller than any classical length characterizing the 3-geometry.
On the quantum level, once the classical phase space functions are mapped into operators, the well-known more difficult issues of quantum fluctuations, non-vanishing expectations
of variances, choices of quantum state, and so on, immediately arise. This is most apparent for the quantum clock \emph{operators} $\hat{T}^{I}$: What does it mean for a non-trivial operator
$\hat{T}^{I}$ to ``take 
the value $\tau ^{I}\in\mathbb{R}$"? In the fully developed quantum regime, of course, there is not even a well-defined 3-geometry at all, so the best one might hope for is that one can find 
some kind of semiclassical regime or limit that supports or approximates a 3-metric like $d(x,y)$.
For now, we will work at a level (classical or semiclassical) where we may safely assume we do have a sufficiently accurate 
$t$-dependent $d(i,j)$ on the lattice, and  discretization error, quantum fluctuations, 
and semiclassical  consistency will be discussed later in section 7 after we see what the Lieb-Robinson bound can tell us about relational framework operator gauge-flow in external time with non-ultralocal constraints. \\
 
 \section{Introduction to Lieb-Robinson Bounds}
 
  Here we provide a brief and hopefully self-contained introduction to Lieb-Robinson bounds. The definitions and theorems will be presented together with some intuition, but we refer the more interested
  reader to references \cite{Hastings} and \cite{Nachtergaele} for the detailed derivations. 
  
  From a pedagogical point of view, it is best to start with the simplest case first: A non-relativistic spin system on a 3-D lattice\cite{Hastings}. So consider the 1-dimensional transverse Ising 
  model for $N$ spins with Hamiltonian 
  \begin{equation}
  H=-J\sum _{i=1}^{N-1} S_{i}^{z} S_{i+1}^{z} + B \sum _{i=1}^{N} S_{i}^{x}. \label{Ising}
  \end{equation}
  This spin Hamiltonian has the form $H=\sum _{Z} H_{Z}$ with $H_{Z}$ supported on $Z$. Lieb-Robinson bounds are most suited to cases where $||H_{Z}||$ decays rapidly with $\mathrm{diam}(Z) >1$.
  Using (\ref{Ising})
  and the metric $d(i,j) =|i-j|$, we see that the Zeeman term has diameter 0 and the Ising (exchange) interaction has diameter one. So $|| H_{Z}||=0$ for $\mathrm{diam}(Z) >1$, and these are 
  examples of ``finite range" interactions. There are also other forms of decaying interactions such as exponential, and so on. One could also place the spins at the vertices of a graph. Then 
  $H$ is again a sum of $H_{Z}$, each $Z$ being two vertices, with $H_{Z}$ non-vanishing only if an edge of the graph links them. In that case the metric $d(i,j)$ could be chosen as the 
  shortest path metric, and gives $||H_{Z}||\ne 0$ only if $\mathrm{diam}(Z) =0, 1$. (This is quite different from the spin networks usually considered in loop quantum gravity!)  \\
  
  When discussing these kinds of spin systems it is natural to give operators the (non-relativistic) time dependence given by Heisenberg evolution:
  \begin{equation}
  \mathscr{O} = \exp [iHt] \mathscr{O}(0) \exp [-iHt], \label{HeisenbEvolutn}
  \end{equation}
  where for simplicity we have taken $H$ to be (explicitly) time independent. Then one has the following \cite{Hastings}:

  Theorem (L-R): Suppose for all sites $i\in\Lambda$ one has the L-R condition:
  
  \begin{equation}
  \sum_{X\owns i} ||H_{X}|| \; |X|\, \exp [ \mu\, \mathrm{diam}(X) ] \le s,	\label{LRCondn}
  \end{equation}
  for some positive real constants $s, \mu$. Let $A_{X}$ and $B_{Y}$ be (bosonic) operators supported on sets $X, Y$, respectively. Then if $d(X, Y) > 0$, one has
   \begin{align}
  ||\, [\,A_{X}(t), B_{Y}(0)\, ] \, ||  & \le 2\, ||A_{X}||\;  ||B_{Y} || \sum _{i\in X} \exp [-\mu\, \mathrm{dist}(i,Y)]\, [\exp (-2s |t| ) - 1] \\
  & \le 2 \, ||A_{X}|| \; ||B_{Y} || \;|X|\, \exp [-\mu\, \mathrm{dist}(X,Y)]\, [\exp (-2s |t| ) - 1].  \label{LRB1} 
\end{align}

The physical interpretation of this bound is made especially lucid by an argument due to Hastings \cite{Hastings}, which we reproduce here because of its later importance:
Given an operator $A$ with support $X$ as above, let $B_{\ell}(X)$ be the ball radius $\ell$ about $X$: $B_{\ell}(X)=\{ i : \mathrm{dist} (i,X) \le \ell\}$. Construct the following
operator:
\begin{equation}
A^{\ell}_{X}(t) = \int \mathrm{d} U \; U A_{X} (t) U^{\dagger}, 
\end{equation}
where one integrates over unitaries $U$ supported on $\Lambda\setminus B_{\ell}(X)$ using the Haar measure.  $A^{\ell}_{X}(t)$ has support $B_{\ell}(X)$. 
Since $U A_{X}(t) U^{\dagger} = A_{X}(t) + U [A_{X}(t), U^{\dagger} ]$, one has
\begin{equation}
\big|\big| \,A_{X}^{\ell}(t) - A_{X}(t)\, \big|\big| \le \int \mathrm{d}U\; \big|\big| \,[A_{X}, U]\,\big |\big|. \label{HastingsBall}
\end{equation}
Using Lieb-Robinson bound (\ref{LRB1}) to bound the integrand on the RHS, we see $A_{X}^{\ell}(t)$ is exponentially operator norm-close to $A_{X}(t)$ provided $\ell$ is sufficiently 
large compared to $2s |t| /\mu$. That is, a time-evolved operator $A_{X}(t)$ may be approximated to exponential accuracy by an operator $A_{X}^{\ell}(t)$ supported on 
$B_{\ell}(X)$. Therefore $B_{\ell}(X)$ has the interpretation of an effective $t$-dependent support for $A_{X}(t)$, and the (norm) ``leakage" of $A_{X}(t)$ out of the ``light-cone"
$B_{\ell}(X)$ is exponentially small.  \\

Most commonly the Lieb-Robinson bound is cast into the following form: Suppose the L-R condition (\ref{LRCondn}) holds, then there is a constant $v_{LR}$ that depends on $s,\mu$ such that for
$\ell =\mathrm{dist}(X,Y)$, and $\ell\ge v_{LR}\, t$,
\begin{equation}
||\, [\,A_{X}(t), B_{Y}(0)\, ]\, || \le \frac{v_{LR}\, |t| \,} {\ell} g(\ell) \,|X|\, ||A_{X}||\, ||B_{Y}||, \label{LRB2}
\end{equation}
and $g(\ell)$ decays exponentially with $\ell$. From the theorem, $v_{LR} = 2s/\mu$. $A_{X}(t)$ can be approximated by $A_{X}^{\ell}(t)$ supported on the set of sites within distance
$\ell = v_{LR} |t|$ of $X$ by an error whose norm is bounded by $\ell ^{-1} v_{LR} \,|t| \,g(\ell) \,|X|\, ||A_{X}||$.  
Bounds on the leakage of information (von Neumann entropy) out of the light-cone were studied
in \cite{Bravyi}. For $H_{Z}$ of finite non-zero range, i.e $||H_{Z}|| = 0$ for $\mathrm{diam}(Z) > R$ for some $R$, the bound may be further improved \cite{Hastings}. If $R=1, ||H_{Z}|| \le J$ then 
one finds $g(\ell)$ decays \emph{faster} than exponentially, roughly $g(\ell ) \sim \exp (-a\ell ^{2})$, for positive constant $a$. However, if $H_{Z}$ has range 0, the discrete equivalent of 
ultralocality, then $\mu$ is undefined since $\mathrm{diam}(X)=0$, and there is more no Lieb-Robinson light-cone. \\

The intuition underlying the exponential decay is the following: From the proof \cite{Hastings} one finds that the $n$-th order term of the exponential comes from a chain 
$H_{Z_{1}}, \dots , H_{Z_{n}}$ such that $Z_{1}\cap X \ne \emptyset , Z_{1}\cap Z_{2} \ne \emptyset , \dots , Z_{n-1}\cap Z_{n} \ne\emptyset , Z_{n}\cap Y \ne\emptyset$, i.e. a chain of $n$
patches $Z_{k}, k=1,\dots , n$ each supporting a local patch of $H$. Successive $H_{k}$ are mutually non-commuting as their ranges overlap, but more distant ones commute as their supports are mutually disjoint. So it is crucial for Lieb-Robinson bounds that $
[H_{Z_{1}}, H_{Z_{2}}]\ne 0$ for $Z_{1}\cap Z_{2}\ne \emptyset$ and $[H_{Z_{1}}, H_{Z_{2}}] = 0$ for $Z_{1}\cap Z_{2} = \emptyset$ for all Hamiltonian patches $Z_{1}, Z_{2}$. This why the (classical) relational framework analog $\{H_{I}(\tau), H_{J}(\tau)\}$ was 
studied earlier in section 3, where the relation between the Lieb-Robinson bound patchy $H_{Z}$ and the relational framework $H_{I}(\tau)$ is given by (\ref{Patchy}). These chains of successively overlapping Hamiltonian patches generate the effective operator support $A^{\ell}_{X} (t)$ (light-cone). \\

While there are clear similarities with some features of relational framework non-ultralocality,
the relational framework Hamiltonians $H_{I}(\tau )$ have a non-trivial (and non-unitary) $\tau(t)$ flow, see(\ref{HPBrac3}).
In particular, the simple 
Heisenberg evolution with a time independent Hamiltonian (\ref{HeisenbEvolutn}) does not apply to $\hat{\mathscr{O}} [f](\tau(t))$, and one requires an Lieb-Robinson bound for $t$ dependent Hamiltonians.  
A non-relativistic Lieb-Robinson bound including this possibility was derived by 
Nachtergaele, Vershynina, and Zagrebnov (NVZ) in 2011  \cite{Nachtergaele}, which will now be sketched. \\

NVZ start with vertices $x\in\Gamma$, where $\Gamma$ is a countable set of vertices. They assume: 
There exists a non-increasing real-valued function $F:[0,\infty) \rightarrow (0,\infty)$ such that 
\begin{align}
||F|| & \doteq \sup _{x\in\Gamma} \sum _{y\in\Gamma} F\big(d(x,y)\big) < \infty , \; \;\mathrm{and} \label{DefF} \\
C & \doteq \sup _{x,y\in\Gamma} \sum _{z\in\Gamma} \frac {F\big(d(x,z)\big) F\big(d(z,y)\big)} {F\big(d(x,y)\big)} < \infty . \label{DefC}
\end{align}
For $\mu >0$ define $F_{\mu}(x) \doteq \exp (-\mu x) F(x)$, so $||F_{\mu}|| < ||F||, C_{\mu} < C$. 
The Hilbert space of states for the subsystem at $x\in\Gamma$ is $\mathscr{H}_{x}$. For finite $\Lambda \subset \Gamma$ the Hilbert space associated with $\Lambda$ is 
$\mathscr{H}_{\Lambda} \doteq \bigotimes  _{x\in\Lambda} \mathscr{H}_{x}$. The algebra of observables supported on $\Lambda$ is $\mathscr{A}_{\Lambda} \doteq 
\bigotimes _{x\in\Lambda} \mathscr{B}(\mathscr{H}_{x})$, where $\mathscr{B}(\mathscr{H}_{x})$ is the set of bounded linear operators on $\mathscr{H}_{x}$. If $\Lambda_{1} \subset \Lambda_{2}$, then identify $\mathscr{A}_{\Lambda_{1}} $ with the sub-algebra $\mathscr{A}_{\Lambda_{1}} \otimes \hat{1}_{\Lambda_{2} \setminus \Lambda_{1} }$ of $\mathscr{A}_{\Lambda_{2}}$, and so $\mathscr{A}_{\Lambda_{1}} \subset \mathscr{A}_{\Lambda_{2} }$. The algebra of local observables is defined as
\begin{equation}
\mathscr{A}^{\mathrm{loc}}_{\Gamma} \doteq \bigcup _{\Lambda \subset\Gamma} \mathscr{A}_{\Lambda}.
\end{equation}
The $C^{*}$-algebra of quasi-local observables $\mathscr{A}$ is the norm completion of $\mathscr{A}^{\mathrm{loc}}_{\Gamma} $.  The support of $A\in \mathscr{A}_{\Lambda}$ 
is the minimal set $X\subset \Lambda$ such that $A = A' \otimes \hat{1}_{\Lambda\setminus  X}$ for some $A' \in \mathscr{A}_{X}$. The generator of the operator dynamics 
is defined for each finite volume $\Lambda \subset \Gamma$ and we confine our interest to Hamiltonian interactions (NVZ were also able to include suitable dissipative terms).
This interaction is such that for each finite $X\subset \Gamma$ and for all $t$, $\Phi (t,X)$ is an operator in $\mathscr{A}_{X}$ and $\Phi ^{*}(t,X) = \Phi (t,X)$. The evolution map
$\mathscr{L}_{\Lambda} (t)$, for any finite $\Lambda\subset\Gamma$ and time $t$, is a bounded linear map $\mathscr{A}_{\Lambda}\rightarrow\mathscr{A}_{\Lambda}$,
\begin{equation}
\mathscr{L}_{\Lambda}(t) (A) \doteq \sum _{Z\subset\Lambda} (i) [ \Phi (t,Z), A] \doteq \sum _{Z\subset\Lambda} \Psi_{Z}(t)(A).
\end{equation}
The $\Psi_{Z}(t)$ are bounded linear maps acting on $\mathscr{A}_{X}$, for any $X\subset\Lambda$ such that $X\supset Z$, which are of the form 
$\Psi_{Z}(t) \otimes \mathrm{id}_{X\setminus Z}$. The $\Psi_{Z}(t)$ have norms that generally depend on $X$, but are uniformly bounded as
$|| \Psi_{Z}(t) || \le 2 ||\Phi (t, Z) ||$. Let $M_{n}=\mathscr{B}(\mathbb{C}^{n})$ be the $n \times n$ complex matrices.
We say a map $\Psi\in\mathscr{B}(\mathscr{A}_{Z})$ is \emph{completely bounded} iff, $\forall n\ge 1$ the linear maps
$\Psi \otimes \mathrm{id}_{M_{n}} $, defined on $\mathscr{A}_{Z} \otimes M_{n}$, are bounded with uniformly bounded norm
\begin{equation}
|| \Psi ||_{\mathrm{cb}} \doteq \sup _{n\ge 1} || \Psi \otimes \mathrm{id}_{M_{n}} || < \infty .
\end{equation}
Specifically, $|| \Psi_{Z} ||_{\mathrm{cb}}$ is a map defined on $\mathscr{A}_{\Lambda}, \forall \Lambda\subset\Gamma$ such that $Z\subset\Lambda $, which is 
\emph{independent} of the choice of $\Lambda$ in $\Gamma$. \\

To obtain an Lieb-Robinson bound, NVZ make the following two hypotheses: Given $\Gamma , d, F$ as above, \\

(1) For all finite $\Lambda\subset\Gamma$, $\mathscr{L}_{\Lambda}$ is norm continuous in $t$, hence uniformly continuous on compact intervals. \\

(2) For each $\Lambda$, there exists $\mu >0$ such that for all $t\in\mathbb{R}$, \\
\begin{equation}
||\Psi || _{t,\mu} \doteq \sup _{s\in [0,t]} \sup _{x,y\in \Lambda} \sum _{Z\owns x,y} \frac {|| \Psi _{Z} ||_{\mathrm{cb}} }{F_{\mu} \big(d(x,y)\big)} < \infty .
\end{equation}
One also finds
\begin{equation}
|| \mathscr{L}_{\Lambda} (t) || \le \sum _{Z\subset\Lambda} ||\Psi_{Z}(t) || \le \sum _{x,y\in\Lambda} \sum _{Z\owns x,y} || \Psi _{Z} (t)||_{\mathrm{cb}} \le
||\Psi ||_{t,\mu} \,|\Lambda |\, || F || \doteq M_{t}.
\end{equation}
By definition of $||\Psi||_{t,\mu}$ one has $M_{s} \le M_{t}$ for $s < t$. \\

Fix some large time $T>0$, and for all $A\in\mathscr{A}_{\Lambda}$ let $A(t)$ for $t\in [0,T]$ be a solution of the ODE
\begin{equation}
\frac{\mathrm{d}} {\mathrm{d}t} A(t) = \mathscr{L}_{\Lambda} \, A(t)\; \mathrm{with}\; A(0) = A. \label{NVZODE}
\end{equation}
Because $|| \mathscr{L}_{\Lambda} (t) || \le M_{T} < \infty $, this ODE has a unique solution defined by $\gamma^{\Lambda}_{t,s}(A)  = A(t)$ for
$0\le s\le t\le T$, where $A(t)$ is the unique solution of (\ref{NVZODE}) for $t\in [s, T]$ with initial condition $A(s) = A$.
We say a linear map $\gamma:\mathscr{A}\rightarrow\mathscr{B}$ for $C^{*}$-algebras $\mathscr{A,B}$ is \emph{completely positive} if the maps
$\gamma \otimes \mathrm{id}_{n}: \mathscr{A} \otimes M_{n}\rightarrow \mathscr{B} \otimes M_{n} $ are positive for all $n\ge 1$. Here positive means positive algebra elements
(i.e. of form $A^{*} A$ ) are mapped to positive algebra elements. NVZ showed that the map $\gamma _{t,s}$ is a unit preserving, completely positive map.
For the one parameter group of automorphisms induced by the Hamiltonian generators $\Phi( t,Z)$, the NVZ version of the Lieb-Robinson bound states:  \\

There are constants $v,\mu, c$ such that for $A\in\mathscr{A}_{X}, B\in\mathscr{A}_{Y}$,
\begin{equation}
|| \,[ A,  B(t) ] \, || \le C(A,B) \exp \bigg[-\mu \big(d(X,Y) - v t\big)\bigg], \label{NVZ_CVE}
\end{equation}
where $C(A,B) = c\, || A ||\, ||B ||\, \min \big(|X|,|Y|\big)$.  More specifically, given assumptions (1) and (2) above,  NVZ's theorem 2 states that
\begin{align}
|| \,[ A(s), B(t) ]\, || \le\, & ({2}/{C_{\mu}}) \, ||A||\, ||B||\, ||F|| \min \big(|X|,|Y|\big) \times\nonumber\\
& \times\exp \big(-\mu\, d(X,Y)\big) \Big[ \exp \big(||\Psi ||_{t,\mu} \, C_{\mu} (t-s) \big) -1\Big], \label{NVZ_LRB}
\end{align}
for $X,Y \subset\Lambda, X\cap Y = \emptyset$. Notice the bound is \emph{uniform} over the chosen $\Lambda$. To extend  to uniformity over $\Gamma$,  the definition of  $||\Psi || _{t,\mu}$
in assumption  (2) above should have the sup over $x,y$ altered from $\Lambda$ to $\Gamma$.\\

In order to adapt this result to relational framework gauge flow, we need to know that the derivation introduces the quantity
\begin{equation}
C_{B}(X, t) \doteq \sup _{\mathscr{T}\in \mathscr{B}_{X}} \frac {|| \mathscr{T} \gamma ^{\Lambda} _{t,s} (B) ||} {|| \mathscr{T} ||_{\mathrm{cb}} },
\end{equation}
where for $X\subset\Lambda$, $\mathscr{B} _{X}$ is the subspace of $\mathscr{B}(\mathscr{A}_X)$ of completely bounded linear maps vanishing on the identity.
Here NVZ use the cb-norm (in contrast to the standard norm) to make the denominator independent of $\Lambda\subset\Gamma$.
NVZ's derivation gives
\begin{align}
C_{B}(X, t)  & \le C_{B}(X,s) + \sum _{Z\cap X\ne\emptyset} \int _{s}^{t} || \mathscr{L}_{Z}(r) || \,C_{B}(Z, r)\, \mathrm{dr}, \; \text{with}\; \label{NVZ_A}\\
|| \mathscr{L}_{Z}(t) || & = || \Psi _{Z}(t) || \le ||\Psi ||_{t, \mu } \sum _{x,y\in Z} F_{\mu } \big(d(x,y)\big). \label{NVZ_B} 
\end{align}
One also has that $C_B(Z,s)= || B ||$ if $Z\cap Y \ne\emptyset $ and otherwise vanishes. Iterating these equations produces a Dyson expansion:
\begin{align}
C_{B}(X, t)  & \le || B || \sum _{n=1}^{\infty} \frac{1}{n!} (t-s)^{n}\, a_{n}, \label{NVZ_C}\\
a_{n} & = \Big( || \Psi ||_{t,\mu} \Big) ^{n} \big( C_{\mu }\big)^{n-1} \sum _{x\in X, y\in Y} F_{\mu} \big(d(x,y)\big). \label{NVZ_D} 
\end{align}
This implies equations (\ref{NVZ_CVE}) and (\ref{NVZ_LRB})  above with the (non-relativistic) spatially uniform Lieb-Robinson velocity
\begin{equation}
v_{LR} = ||\Psi ||_{t,\mu} \frac{C_{\mu}}{\mu} \le  ||\Psi ||_{T,\mu} \frac{C_{\mu}} {\mu} , \label{NVZ_VLR}
\end{equation}
for all $t\in[0,T]$ (temporally uniform bound). This bound on $v_{LR}$ can be utilized to bound the norm-leakage of operators outside the Lieb-Robinson light-cone,
analogously to what was performed earlier for the time-independent Hamiltonian case. \\
 
 \section{Lieb-Robinson Bounds for the Discretized Relational Framework}
  
We now apply the NVZ version of a Lieb-Robinson bound \cite{Nachtergaele} to the relational framework discretized as previously described on some lattice or network.  
It is essential to handle appropriately the fact that the 3-metric $d(x,y)$ is both slicing $\{\tau(s)\}$ and slice $s$ (external time) dependent, and the Lieb-Robinson bound should 
preserve the necessary gauge invariance. Strictly speaking, $d(x,y)$ should then be denoted as $d(x,y; \{\tau (s)\}, s)$ for the slice at external time $s$ in slicing $\{\tau(s)\}$,
however we will continue to use the abbreviated form $d(x,y)$ for convenience. The reader should bear in mind the suppressed slicing and slice dependence. \\

The key initial step is to replace NVZ's $\Phi (t,  Z)$ by 
\begin{equation}
\Phi (t, Z) \rightarrow H_{Z}(t) =\sum _{i=0}^{N_{c}-1} \Big( \frac{\partial\tau ^{j_{c}(Z)}_{i}(t)} {\partial t} \Big )\, H^{j_{c}(Z)}_{i} (\tau (t) ), \label{NVZ_F}
\end{equation}
from the spatially discretized version of (\ref{Patchy}). Equation (\ref{NVZ_A}) may be iterated as $C_{B}(X,t)\le \sum _{n=1}^{\infty} C_{B}^{(n)} (X,t)$, where the $n=0$ iterate
vanishes since $X\cap Y = \emptyset $. For simplicity we first focus on the $n=1$ term:
\begin{align}
C_{B}^{(1)} (X,t) & \doteq \sum _{Z\cap X\ne\emptyset} \int _{s}^{t} || \Psi_{Z} (r) || \, C_{B} (Z,s)\, \mathrm{dr}  \label{NVZ_G} \\
& \le 2\, || B || (t-s) \sum _{j\in X} \sum _{Z\owns j, Z\cap Y\ne\emptyset} \sup _{\tilde{s}\in [s,t]} || H_{Z} (\tilde{s}) ||. \label{NVZ_H} 
\end{align}
The factor 2 on the RHS comes from $|| \Psi_{Z}(r) || \le 2\, || H_{Z}(r) ||$, where 2 enters from bounding $|| [H_{Z}(r), A] || $ by $ 2\, || H_{Z}(r) || \, ||A||$.
The general idea of the rest of the derivation is to insert strategically placed uniforming bounds (sups) after introducing an appropriate factor of $F_{\mu}$. The ranges of the sups
 are also important and have to be selected with care. We also use $\sup (AB) \le \sup(A) \sup (B)$ for $A,B >0$.
  \begin{equation}
 C_{B}^{(1)} (X,t) \le 2\, || B ||\, (t-s) \sum_{j\in X, k\in Y} \sum _{Z\owns j,k} \sup _{\tilde{s}\in [s,t]} \Bigg[ \frac{|| H_{Z}(\tilde{s}) || } {F_{\mu}(d(j,k))} F_{\mu}(d(j,k))\Bigg],
 \end{equation}
 and using the RHS of (\ref{NVZ_F}), 
  \begin{align}
C_{B}^{(1)} (X,t) & \le 2\,|| B || \, (t-s) \sum _{j\in X, k\in Y} \sum _{Z\owns j,k} \big(\sup _{\tilde{s}\in [s,t]} \big) \big(\sup _{i_{2}\in[0,N_{c}-1]} \big) \Bigg[ N_{c} \Big| \frac {\partial
 \tau _{i_{2}}^{j_{c}(Z)} (\tilde{s})} {\partial\tilde{s}} \Big| \Bigg] \times \nonumber \\
 & \times \Bigg(\big(\sup _{\tau(\tilde{s}) }\big) \big( \sup _{i_{1}\in [0,N_{c}-1]}\big) \big(\sup _{\tilde{s}\in [0,T]}\big) \Bigg[\frac{|| H^{j_{c}(Z)}_{i_{1}} (\tau (\tilde{s}) )||  } {F_{\mu} 
 \big(d(j,k)\big)} \Bigg] \Bigg)\times \nonumber \\
 & \times\Bigg( \big(\sup _{\tilde{s}\in [s,t]} \big) F_{\mu} \big(d(j,k)\big) \Bigg), 
 \end{align}
 where a bounding sup over slicings $\{\tau(\tilde{s})\}$ has been inserted into the middle factor.
 Henceforth we will take $\partial\tau _{i}^{j}(t)/\partial t \ge 0$, for all  $t, i, j$, which is just the monotonicity condition necessary for freedom from anomalies discussed earlier, so the 
 absolute values in the first line may be omitted. We have also ``extended" the sup over $\tilde{s}$ in the factor containing $|| H ||$ 
 from $\tilde{s}\in [s,t] $ to $\tilde{s} \in [0,T]$, with 
 $0\le s < t\le T$, where recall $T$ is some ``large" external time. 
 
Next,  first  bound the first factor, containing the $\tilde{s}$ derivatives of $\tau$, by taking an overall sup over $j_{c} \in \Gamma$, 
thereby rendering that factor independent of $Z$.
Returning then to the middle factor with $|| H ||$, bring in the sum over $Z$, and expand $j\in X, k\in Y$ to $x,y \in\Lambda\supset Z$, for some chosen $\Lambda\subset\Gamma$ of finite
cardinality.
By expanding  $j,k$ to $x,y \in \Lambda$,  more positive terms were added.
Hence the sup factor containing $|| H^{j}_{i_{1}}(\tau ) ||$ may be bounded by
 \begin{equation}
 \big(\sup _{\tau(\tilde{s})} \big) \big(\sup _{i\in[0,N_c -1]}\big) \big(\sup _{\tilde{s}\in [0,T] } \big) \big(\sup _{x,y\in \Lambda} \big) \sum_{Z\owns x,y}
 \Bigg[ \frac{|| H^{j_{c}(Z)}_{i} (\tau (\tilde{s}) )||  } {F_{\mu} \big(d(x,y)\big)} \Bigg] ,
 \end{equation}
which is now conveniently independent of $j,k$.  Therefore the $j, k$ sum on the far left may then be moved all the way to the right to act only on the third factor, 
containing  only $F_{\mu}(d(j,k)).$
Thus,
\begin{align}
C_{B}^{(1)} (X,t) & \le 2\, || B ||\, (t-s)  \Bigg( \big(\sup _{j_{c}\in\Gamma} \big) \big( \sup _{i\in[0,N_C - 1]} \big) \big( \sup _{\tilde{s}\in [s,t]} \big) \Bigg[ N_{c} 
\frac {\partial\tau ^{j_{c}}_{i} (\tilde{s}) } {\partial\tilde{s}} \Bigg] \Bigg) \times\nonumber \\
& \times \Bigg(  \big(\sup _{\tau(\tilde{s})} \big) \big(\sup _{i\in[0,N_C -1]}\big) \big(\sup _{\tilde{s}\in [0,T] } \big) \big(\sup _{x,y\in \Lambda} \big) 
 \sum _{Z\owns x,y}  \frac{|| H^{j_{c}(Z)}_{i} (\tau (\tilde{s}) )|| } {F_{\mu} \big(d(x,y)\big)} \Bigg) \times\nonumber\\
& \times \Bigg( \sum _{j\in X, k\in Y}\big(\sup _{\tilde{s}\in [s,t]} \big) 
 F_{\mu}\big(d(j,k)\big) \Bigg). \label{FUN} 
  \end{align}
 
This may be written more compactly as
\begin{align}
C_{B}^{(1)} (X,t) \le & 2\, || B ||\, (t-s) \, I_{X,Y}(F_{\mu})\, \Bigg[ \big(\sup _{j_{c}\in\Gamma} \big) \big( \sup _{i\in[0,N_c -1]}\big) \big(\sup_{\tilde{s}\in [s,t]}\big) \Bigg( N_{c} \frac {\partial\tau^{j_{c}}_{i} (\tilde{s}) }
{\partial\tilde{s}} \Bigg) \Bigg]\times\nonumber\\
& \times || H ||_{T,\mu}\, , \label{NVZ_L}
\end{align}
where we have set
\begin{equation}
I_{X,Y} (F_{\mu}) \doteq \sum _{j\in X} \sum _{k\in Y} \big(\sup _{\tilde{s}\in [s,t]} \big) F_{\mu}\Big(d(j,k)\Big|_{\tau(\tilde{s}),\tilde{s}}\Big).
\end{equation}
Similar to NVZ, we assume there exists a real $\mu >0$ such that
\begin{equation}
|| H ||_{T,\mu} \doteq \big(\sup _{\tau(\tilde{s})} \big) \big(\sup _{i\in[0,N_C - 1]}\big) \big(\sup _{\tilde{s}\in [0, T]}\big) \big(\sup _{x,y\in\Lambda}\big) \sum _{Z\owns x,y} 
 \frac{|| H^{j_{c}(Z)}_{i} (\tau (\tilde{s}) )|| } {F_{\mu} \big(d(x,y)\big)} < \infty . \label{HTmuDef}
 \end{equation}
 From (\ref{HTmuDef}) we see $|| H ||_{T,\mu}$ is a bound  temporally uniform over $\tilde s\in[0,T]$ and spatially uniform over the chosen finite subset $\Lambda\subset\Gamma$.  
 This occurs because $\Lambda\subset\Gamma$ is \emph{any} finite lattice containing $Z\subset\Lambda$, and $|| H^{j_{c}(Z)}_{k} (\tau (\tilde{s}) )||$ is independent of the choice
 of $\Lambda$ within $\Gamma$. One may extend the definition of $|| H ||_{T,\mu}$ to be spatially uniform over countable $\Gamma$ as in  NVZ, by changing the sup over $x,y$ 
 in (\ref{HTmuDef}) from $\Lambda$ to $\Gamma$.
 \\
 
 We now examine $I_{X,Y} (F_{\mu})$ in some more detail. Recall $F_{\mu}(x) \doteq \exp (-\mu\, x) F(x)$ is a positive real-valued, non-increasing function of its non-negative real argument, thus for a fixed slicing
 \begin{equation}
\sup _{\tilde{s}\in [s,t]} F_{\mu} \big(d(x,y)\big) \le F_{\mu} \Big( \inf _{\tilde{s}\in [s,t]} d(x,y) \Big), 
\end {equation}
and
\begin{equation}
 I_{X,Y} (F_{\mu}) \le \sum _{x\in X} \sum _{y\in Y} F_{\mu} \Big( \inf_{\tilde{s}\in [s,t]} d(x,y)\Big).
 \end{equation}
 Alternatively, in the discretized model under study, $X,Y$ are both finite sets, so
 \begin{align}
I_{X,Y} (F_{\mu}) & = \sup _{\tilde{s}\in [s,t]} \sum _{x\in X} \sum _{y\in Y} F_{\mu} \big(d(x,y)\big) \nonumber \\
 & \le \sup _{\tilde{s}\in [s,t]}  \min \big(|X|,|Y|\big) \sup _{y\in m(X,Y)} \sum _{x\in M(X,Y)} F_{\mu} \big(d(x,y)\big), 
 \end{align}
where $m(X,Y) \doteq Y$ and $M(X,Y) \doteq X$ if $ |X|>|Y|$, otherwise $m(X,Y) \doteq X$ and $M(X,Y) \doteq Y$. Since $F_\mu$ is positive,
this may be  bounded by ``expanding" both $m(X,Y)$ and $M(X,Y)$ to $\Gamma$ to yield
\begin{equation}
 I_{X,Y} (F_{\mu}) 
  \le \min \big(|X|, |Y|\big) \exp \big[ -\mu\inf _{\tilde{s}\in [s,t]} d(X,Y)\big] \sup _{\tilde{s}\in [0,T]} ||F||,
 \end{equation}
 where as a reminder, $||F|| \doteq \sup _{x\in\Gamma}\sum_{y\in\Gamma} F\big(d(x,y)\big) < \infty $ contains an implicit $\tilde{s}$ dependence through $d(x,y)$. \\
 
 When one bounds the higher order terms $n>1$ in the Dyson expansion (\ref{NVZ_C}) for $C_{B}(X,t)$, at order $n$ one initially inserts $n-1$ factors of
 \begin{equation}
 \frac {F_{\mu}\big(d(x,z)\big) F_{\mu}\big(d(z,y)\big)} {F_{\mu}\big(d(x,y)\big)} \le C\, \frac{\exp \big[-\mu \big(d(x,z)+d(z,y)\big) \big] } {\exp \big[-\mu\, d(x,y) \big] } < C,
 \end{equation}
where the triangle property of the 3-metric on a fixed slice of a fixed slicing has been applied. Recall the positive real constant $C$ is defined by (\ref{DefC}), and $F_{\mu}(x)=e^{-\mu x} F(x)$.
There are also $n$ factors of $H_{Z}$. Expanding $ C_{B}(X,t) \le ||B|| \sum _{n=1} \hat{a}_{n} (t-s)^{n}/n!$, one bounds
\begin{align}
\hat{a}_{n} \le & \big(2\,||H||_{T,\mu}\big)^{n} \Big[\big(\sup _{\tau(\tilde{s})} \big) \big(\sup _{\tilde{s}\in [0,T]}\big) \, C_{\mu}\big(d(\tau(\tilde{s}),\tilde{s})\big) \Big]^{n-1} \times\nonumber \\
& \times \Bigg[ \big(\sup _{j_{c}\in\Gamma}\big)\big(\sup _{i\in[0,N_c -1]} \big) \big(\sup _{\tilde{s}\in [s,t]}\big) \Bigg(N_{c} \frac{\partial\tau^{j_{c}}_{i}(\tilde{s})} {\partial\tilde{s}}\Bigg)\Bigg]^{n} \times\nonumber \\
& \times \Big[\big(\sup _{\tilde{s}\in [s,t]}\big) \sum _{j\in X} \sum _{k\in Y} F_{\mu}\big( d(j,k)\big|_{\tau(\tilde{s}),\tilde{s}}\big) \Big]. 
\end{align}
We now assemble all these intermediate steps into the final result. Define 
\begin{equation}
\tilde{C}_{\mu} \doteq \big(\sup _{\tau(\tilde{s})}\big) \big(\sup _{\tilde{s}\in [0,T]}\big) C_{\mu}.
\end{equation}
Because external time re-parameterization (1-diff) invariance  will require an Lieb-Robinson bound restricted to infinitesimal time increments $(t-s)\rightarrow 0$, we set $\delta t \doteq (t-s) \rightarrow 0$ in the above expressions. Also define
\begin{align}
&  \big(\sup _{\tilde{s}\in [t, t+\delta t]}\big) ||F\big(d(\tau(\tilde{s}), \tilde{s}) ||  \le \big(\sup _{\tau(\tilde{s})}\big) \big(\sup _{\tilde{s}\in [0, T]}\big) \big(\sup _{x\in
\Gamma}\big) \sum _{y\in\Gamma} F\big(d(x,y;\tau(\tilde{s}), \tilde{s})\big) \nonumber \\
& \le \sup _{x\in\Gamma} \sum _{y\in\Gamma} F\big(\inf _{\tau(\tilde{s})} \inf _{\tilde{s}\in [0,T]} d(x,y)\big) \doteq ||\tilde{F}||, 
\end{align}
and $|| \tilde{F}|| $ is slicing and slice independent. Then one has the relational framework-Lieb-Robinson bound, 
\begin{align}
\big|\big|\, [A_{X}(t), B_{Y}(t+\delta t)]\,\big|\big|\bigg|_{\tau(s)} \le & \, \frac{2}{\tilde{C}_{\mu}}\, ||A||\, ||B||\, ||\tilde{F}|| \min \big(|X|,|Y|\big) \times\nonumber\\ 
& \times\Bigg[ \exp \Big( -\mu\,\mathscr{E}\big(\tau(t)\big) \Big) - 
\exp \Big( -\mu\inf _{s\in [t,t+\delta t]} d(X,Y)\Big) \Bigg],
\label{RFLRB}
\end{align}
where the $|_{\tau(s)}$ on the LHS indicates one is referring to a single slice at some $t\in[0,T]$ within an arbitrary slicing $\tau(s)$. As a reminder, $X, Y$ are the supports of $A(t),
B(t)$, respectively, with $X\cap Y= \emptyset$. We have denoted
\begin{align} 
\mathscr{E}\big(\tau(t)\big) \doteq & \Bigg[\inf _{s\in [t,t+\delta t]} d(X,Y;\tau(s),s)\Bigg] - \Bigg(\frac{2\tilde{C}_{\mu}}{\mu} N_{c}\, || H ||_{T,\mu}\Bigg) (\delta t)\times\nonumber\\
& \times \Bigg[\big(\sup _{j\in\Gamma}\big) \big(\sup _{i\in[0,N_c -1]}\big) \big( \sup _{s\in [t,t+\delta t]} \big)
\bigg(\frac{\partial\tau^{j}_{i}(s)}
{\partial s} \bigg)\Bigg]. \label{DefEsc} 
\end{align}
We will refer to the quantity
\begin{equation}
v_{LR} \doteq \frac {2\tilde{C}_{\mu} } {\mu} \,N_{c}\, || H ||_{T,\mu} \label{RFVLR}
\end{equation}
as the relational framework Lieb-Robinson velocity. \\

\section{Invariance and Other Properties of the relational framework Lieb-Robinson Bound}

The relational framework Lieb-Robinson bound (\ref{RFLRB}) has a LHS that refers to the $t$-differential $(\delta t)$ behavior of the norm of an operator commutator between discretized 
observables $A,B$ near a single slice of some 
arbitrary slicing. Typically we take the operators $A,B$ to have the relational framework form $\Delta\big(\hat{\mathscr{O}} [f] (\tau )\big)$, so they are discretized Dirac observables. 
The RHS has many factors, some of which are uniformly bounded over slicings and slices, and others that are slicing and/or slice dependent. We now describe those dependencies.  
Recall that a change of slicing is a gauge transformation, and a choice of slice
$s$ within a slicing is a gauge fixing. By definition (\ref {HTmuDef}), $\mu \sim 1/\mathrm{diam} (Z)$ is chosen so that $|| H ||_{T,\mu}$ is finite and includes sups over all slicings $\tau(\tilde{s})$
and over all slices at $\tilde{s}$ within those slicings. Therefore, in addition to being spatially and temporally uniform, $\mu$ and $|| H ||_{T,\mu}$ are slicing and slice uniform as well. 
Moreover, $\tilde{C}_
{\mu} \doteq \sup _{\tau(s)} \sup _{s\in [0, T]} C_{\mu}$ is $\mu$ and $F$ dependent, and hence slicing and slice uniform too. Thus $(2/\mu)\, \tilde{C}_{\mu}\, || H ||_{T,\mu}$ and then $v_{LR}$
are both slicing, slice, and spatio-temporally uniform. The same conclusion holds for the prefactor of the exponentials on the RHS of (\ref{RFLRB}). \\

By comparison with the simple $t$ independent Hamiltonian non-relativistic  Lieb-Robinson bound (\ref{LRB1}), we see the exponentially damped leakage from a local light-cone is governed by the 
quantity $\mathscr{E}$ in (\ref{DefEsc}) above. So when
\begin{equation}
\delta D(t) \doteq \inf _{s\in [t,t+\delta t]} \Big( d(X,Y;\tau(s), s) \Big) > v_{LR} \, \sup _{s\in [t,t+\delta t]} \Big[ \frac {\partial\tau ^{j}_{k} (s)} {\partial s} \,\delta t  \Big] 
\label{dDDef}
\end{equation}
there is exponentially small leakage of the operator norm from the local light-cone during $\delta t$. $\delta D(t)$ is a slicing dependent and slice dependent (slice near $t$) quantity. It is also
$t$ reparameterization (1-diff) invariant under $t\rightarrow t' \doteq f(t), \delta t \rightarrow \delta t' =  f'(t)\, \delta t $ with $f'(t) > 0$ for $t\in [0,T]$,  treating $\tau ^{I}$ and $H_{I}(\tau )$ as 
$(3+1)$ scalars, and noting that such a relabelling of the slice from $t$ to $t'$ does not affect the 3-metric $d$ on the slice. At a fixed slicing and slice, $\delta D(t)$
is also 3-diff invariant because the 3-metric $d$ is, even though the spatial discretization (lattice) itself is not  3-diff invariant.  The factor $\sup _{s} (\partial\tau /\partial s)\, \delta t$ on the RHS of 
(\ref{dDDef}) has the same properties. Hence one has on-shell (3+1)-diff invariance of the relational framework Lieb-Robinson bound local light-cone. \\

As shown by Dittrich \cite{Dittrich_RF}\cite{Dittrich_GR}, on-shell one can classically embed every slicing into a 4-manifold with a Lorentzian 4-metric.
The tangent bundle of this 4-manifold may also be smoothly partitioned in a 
$(3+1)$-diff invariant way using that Lorentzian metric to define local null directions and so generate a 4-metric based null-cone. 
The relational framework-Lieb-Robinson bound local light-cone should coincide with or bound the 4-metric null-cone,  but this has 
not yet been explicitly established. The 4-metric null-cone, however, does not address the important issue of observable commutator leakage outside
the light cone, which is the crux of the relational framework-Lieb-Robinson bound. \\

One can also derive a relational framework-Lieb-Robinson bound local light-cone structure with $v'_{LR} \sup _{s} (\partial\tau / \partial t) \delta t$ replacing $v_{LR} $ $\sup _{s} (\partial\tau / \partial t) \delta t$ in a new 
$\delta \tilde{D}(t)$ which is \emph 
{slicing independent} (containing a $\sup _{\tau (s)}$) but is \emph {slice dependent} (still retaining the $\sup _{s\in[t,t+\delta t]} $). That is, one fixes some external $t$ parametrization $\tau^{I}(t)$, and looks at slices within $\delta t$ of $t$ \emph {over all} the slicings $\tau(t)$. But this construction seems less physically natural than the one described above, where $\delta D(t)$ is both slicing and slice dependent, so we will not discuss it further. \\

A differential Lieb-Robinson bound local light-cone can be ``integrated forwards"  in $t$ from slice to slice within a single slicing to generate ``support tubes" for observables. 
To do this, one constructs operators $A^{\ell}_{X} (t, t+\delta t)$ over the external 
time interval $[t,t+\delta t]$ by the Hastings method described in section 4 which are exponentially accurate $t$-dependent supports 
for a discretized Dirac observable gauge-evolving in external time from $t$
to $t+\delta t$. This then may be iterated for succeeding slices spaced by intervals $\delta t \rightarrow 0$. This is precisely analogous to how one ``integrates" null-cones on a curved Lorentzian manifold to generate causal curves from the locally Minkowskian geometry. \\

\section{Physical and Conceptual Questions}

We have explored non-ultralocal constraints with the relational framework and derived an external time differential local light-cone structure based on Lieb-Robinson bounds using a discrete spatial lattice or network model. 
Several physical and conceptual issues about this spring to mind and we discuss those here.  This discussion is by necessity less mathematically rigorous and considerably speculative in some cases.  \\

Question 1: Aside from slicing and slice dependent factors like $\Big(\partial\tau/\partial s\Big)\delta t$, are $v_{LR}$ and the local light-cone ``the same" as the classical spacetime into which
 the relational framework is embedded varies? That is, basically one is taking sups over a large spacetime to construct $v_{LR} \sup _{s} (\partial\tau /\partial s) \delta t$. But what happens when the entire
 spacetime, initial conditions and so on, are altered? Would $\mu$ stay the same? If $\mu$ changed, then according to (\ref{RFVLR}) $v_{LR}$ would also change as the 
 spacetime under investigation was altered, a potentially fatal physical pathology if we expect $v_{LR}$ to be (or bound) the speed of light. \\
 
 One way to avoid such an early demise for the relational framework Lieb-Robinson bound would be that 
 $1/\mu \sim \mathrm{diam} (Z)$, the proper ``typical size" of single on-shell Hamiltonian patches, is a proper length much smaller than the minimum over the classical spacetimes of any 
 proper curvature scales $L_{c} (x)$ they contain. That is, one could interpret $1/\mu$ as some kind of proper finite range $\xi$ of the on-shell constraints or Hamiltonians.
 Thus, if $\xi$ is a microscopic
 scale relative to classical geometrical scales: $\xi \ll \min _{x} L_{c}(x)$ uniformly for all the classical spacetimes under consideration,
 then different classical geometries but with the \emph{same constraints} will have the same $\mu ,||H||_{T,\mu},\tilde{C}_{\mu}$ 
 and thus $v_{LR}$. This takes the constraints to be nonlocal, homogeneous in form, and uniformly bounded; it
 essentially requires a large separation of physical scales, which is common throughout physics.
 In addition, $\xi$ would also have to be far smaller than any particle physics lengths that have be probed so far for there 
 not to have been any evidence yet of nonlocality. But this does not mean that $\xi$ has to be on the order of the Planck scale $L_P\sim 10^{-35}$ m, but it would 
 certainly require $\xi << 2\cdot 10^{-19}$\,m (1 TeV). $\xi$ would still have to be longer than the scale needed to have a well defined 3-metric on each slice, which at least requires that 
 $\xi\gg L_P$. \\
 
 Question 2: What kind of terms in $H_{I}(\tau)$ would generate the required non-ultralocality? Specifically, 
 would gradient terms (of any order) suffice to produce the non-ultralocality for a Lieb-Robinson bound?\\
 
 This question is closely tied to discretization. In a typical discretization one replaces $\nabla_{x} \psi (x)  \rightarrow (\psi (j+1)- \psi (j))/b$, where
 $b$ is some kind of lattice constant, and $j$ is a site index corresponding to the continuum coordinate $x$. 
 So $H_{I}$ containing a point-wise gradient in the continuum, such as $\psi (x) \nabla\psi (x) $, becomes nonlocal in the discretization since it would couple sites $j$ and $j+1$. But
 in the continuum $H_{I}$ remains firmly ultralocal. Indeed, all the well known actions for continuum canonical gravity such as the 3+1 ADM decomposition or the Holst action are ultralocal,
 and for those cases $\xi=0$, and then there is no more Lieb-Robinson local light-cone. \\
 
 The answer to the question is no, continuum gradients alone are insufficient for non-ultralocality. The reason for the negative answer is any discretization perceives a continuum gradient
  as a lattice constant dependent contribution to $\xi(b)=1/\mu(b)$.
 An $n$-th order gradient is computed by discretization to give a contribution
 $\xi\propto nb$, which vanishes in the continuum limit $b\rightarrow 0$. Gradients have no non-zero natural scale
  in the continuum. Hence for a local light-cone to emerge by Lieb-Robinson bounds, the non-ultralocal $H_{I}$
 cannot be constructed from products fields and their gradients at a single point. Both $\xi(b)$ and $||H_{I}||_{T,\mu }$ must be independent of $b$ as $b\rightarrow 0$. Physical
 quantities like $v_{LR}$ and the local light-cone cannot depend on any cutoff scale like $b$. This is the lesson of the renormalization group for background  dependent quantum field theory, 
 and also requires discretization independence 
 in the continuum limit for gravitational theories. In fact, as found in \cite{Dittrich_etal}, discretization independence implies nonlocality in 4D discrete quantum gravity. \\

 This question and its answer lead us to ask ... \\
 
 Question 3: What is the role of the discretization in the Lieb-Robinson bound? \\
 
 Contact with physical reality occurs when $b\rightarrow 0$ because any non-zero $b$ spatial discretization by itself breaks 3-diff invariance since 
 the lattice or network is not generally mapped into itself by a 3-diff \cite{Dittrich_Diff}.  There are, however, perfect actions for discretizations that do recover the 
 requisite invariance as $b\rightarrow 0$ for several
 models \cite{Bahr_Dittrich}. Even for these special cases, the discretization will coarse-grain or voxelate the metric information at least over the scale $b$. This classical error makes the 
 discretized $d(j,k)$ acquire a $b$ dependence. Again we have to assume that $\lim_{b\rightarrow 0} d(x,y;b) $ is well-defined so $v_{LR} \sup_{s} (\partial\tau /\partial s) \delta t$ and $\delta 
 D(t)$ are all also well-defined in that continuum limit, where the classical voxelation error disappears. \\
 
 Question 4: How might the required non-ultralocality of $h_{I}, H_{I}(\tau)$ arise? That is, if one starts from some bare classical constraints specified by $H^{\mathrm{cl}}_{I}$ that are 
 ultralocal, how can one end up with effective quantum constraints generated by a non-ultralocal $H^{\mathrm{eff}}_{I}$? \\
 
 The simple answer is that the real quantitative origins of non-ultralocality $h_{I}$ lie beyond the scope of the present work. Indeed, this is like asking what atomic physics lies behind the 
 exchange coupling constant $J$ in the Ising model (\ref{Ising}).
 The quantum algebra $\mathfrak{U}$ at the kinematic level is generated by $\hat{Q}_{a}(\tau ),\hat{P}_{a}(\tau )$ through their canonical commutation relations. 
 From that algebra we can find its representation carrier Hilbert space by the GNS construction from any positive linear functional (state) $\rho$ on $\mathfrak{U}$. The quantum constraints are
 $\rho [\hat{P}_{I} + \hat{h}^{\mathrm{eff}}_{I}] =0$, i.e. the quantum fluctuations satisfy the effective constraints in the mean (as an expectation value). A similar process occurs in 
 background-dependent quantum field theory where quantum corrections due to loops are taken into account using a field dependent effective (``dressed") action $\Gamma[\Phi]$ rather than the bare classical
 action $I[\Phi ]$ \cite{Weinberg_EffActn}.  Connected vacuum-vacuum quantum field theory amplitudes can be computed using tree-level (mean field) Feynman graphs
 with vertices using  $\Gamma[\Phi]$ instead of $I[\Phi ]$. A similar
 ``dressing" by quantum fluctuations might render $\hat{H}^{\mathrm{eff}}$ non-ultralocal. Unfortunately the quantitative details are out of present computational reach for background independent quantum gravity without the well-defined path integral technology of quantum field theory.\\
 
 A more speculative answer is that we may not know the ``true" physical action or constraints  for gravity at very short but not yet Planckian lengths, only that they look local 
 as far we can tell from our experience at long length scales (above $2\cdot  10^{-19}$\,m or energies up to 1 TeV). 
 In that case, the Lieb-Robinson local light-cone would be a long length scale manifestation of non-ultralocality of those otherwise inaccessible short length scale constraints,
 a hint that we might not be aware of some deeper physics. One possibility is that non-commutative products
such as the Moyal-Weyl-Groenewold $\star$-product and deformed 
diffeomorphisms could play a role. While non-commutative field theories still have Lagrangian densities and Noether currents, the products of objects in the Lagrangian
density are nonlocal \cite{Schenkel}. The scale of that nonlocality might be associated with $\xi$ above. 
However, non-commutative geometry is not a quantization of  the underlying manifold in the sense that it does not promote phase space functions to operators. 
So non-commutative manifolds do not describe quantum fluctuations of geometry, and they are on-shell descriptions. Instead they require that the geometric and matter field actions be 
invariant under deformed diffeomorphisms of the non-commutative manifold, which are nearly the standard diffeomorphisms normally used to describe manifolds 
like $\Sigma$ that leave the usual local
action invariant. At length scales larger than the nonlocal effects, the noncommutative symmetries and constraints should approach the standard ones 
while an on-shell Lieb-Robinson light cone emerges from
a relational framework based on the non-commutative action. It is worth noting in this regard that the nonlocality induced by non-commutative geometry is not of the gradient (spurious) type. \\
 
 Question 5: In the relational framework-Lieb-Robinson bound which variables are quantum and which are classical? \\
 
 The operators $\hat{A}, \hat{B}$ appearing in the Lieb-Robinson bound analysis include those constructed from $\hat{Q}_{a}(\tau ),\hat{P}_{a}(\tau )$, 
 which may describe either geometrical or possibly matter degrees of freedom.  These are treated fully quantum mechanically from the on-shell gauge-flow (\ref{GEvolutn}). The $\tau_{I}$
 are real parameters and are never promoted to quantum operators, so they do not acquire fluctuations. But the clock variables $T^{I}$ are promoted to operators  $\hat{T}^{I}$,
 making the classical relational framework notion of ''$ T^{I}$ taking the value $\tau _{I}$'' more problematic. Nevertheless,  these clock operators lie buried inside the Hamiltonians $\hat{H}_I$,  
 which are then treated fully quantum mechanically. More crucially, at the heart of the Lieb-Robinson bound lurks $d(x,y)$, the on-slice 3-metric,  computed from geometrical
 variables. In $d(x,y)$ those variables, such as the ADM 3-metric $q_{ab}$, are treated classically, while in $\hat{A}(\tau ), \hat{B}(\tau )$ those same variables are treated 
 quantum mechanically. So there seems to be an inconsistency. That is, inside  $\hat{A}(\tau )$ and $\hat{B}(\tau )$ $\hat{q}_{ab}$ is treated as an element of a non-commutative (quantum) $C^{*}$-algebra, but when $\hat{q}_{ab}$ enters a computation of $d(x,y;\tau(s),s)$ it treated as an element of a commutative (classical) $C^{*}$-algebra. However if we use classical 
 geodesics as a basis for computing
 $d(x,y)$ as an extremum over the quantum expectation $\langle \hat{q}_{ab} \rangle$, then the Lieb-Robinson local light-cone quantity $\delta D(t)$ (see \ref{dDDef})  is mean field 
 (expectation value level) with 
 respect to $ \hat{q}_{ab}$
 over $[t,t+\delta t]$. Within the Hamiltonians $\hat{H}_{I}(\tau)$, $q_{ab}$ is treated as a quantum operator. Thus we can say the Lieb-Robinson local light-cone construction is at least mean field with respect to 
 fluctuations entering  $q_{ab}$, and it must be stabilized against those fluctuations if it is to survive.   \\
 
 Question 6: What parameters, if any, delineate a window of survival for the mean-field Lieb-Robinson local light-cone in its precarious perch among the tensions between the classical and quantum worlds? \\
 
To answer this, we introduce $\delta d$ as the largest rms quantum fluctuation of the the proper lengths $d(i,j)$, and also take the classical discretization error in proper lengths to
be be roughly the same as the lattice constant $b$. $L_{c}$ denotes the shortest classical curvature scale of any 3-geometry under consideration. 
We also have $\xi =1/\mu$, the range (proper patch diameter) of the non-ultralocal constraints. Since a classical metric $d(i, j)$ loses
physical meaning when $d(i, j) < \delta d$, and  quantum fluctuations could play a role in the origin of $\xi$, $\delta d \lesssim\xi$. We can then qualitatively delineate four distinct
physical regimes: \\

(a) $\delta d \lesssim\xi < b \ll L_{c}$ Here the discretization is too coarse to resolve the finite range $\xi$ of the $H_{I}$, which then appears to be ultralocal. The Lieb-Robinson local light-cone
does not emerge. \\

(b) $\delta d<b<\xi\ll L_{c}$ Here the Lieb-Robinson local light-cone is stabilized against the quantum fluctuations $\delta d$, the discretization can resolve the non-ultralocal range $\xi$ of the $H_{I}$, but it does not resolve the quantum fluctuations. The local light-cone is $b$ (discretization) independent once $b\ll\xi$. \\

(c) $b<\delta d<\xi\ll L_{c}$ The proper range $\xi$ is still separated and immune from quantum fluctuations, while the local light-cone is still discretization ($b$) independent. \\

(d) $b<\delta d\simeq\xi \ll L_{c}$ Now $\xi$ is quantum-limited, a fluctuation limited patch size. We still have discretization independence once $b<\delta d$, where the discretization error
is no longer physically relevant. \\

In cases (b) and (c)  the Lieb-Robinson local light-cone will survive provided: (1) the continuum limit $b\rightarrow 0$ is well-behaved,
and (2) $\delta d \ll \delta D(t)$ (see \ref{dDDef}). This means $\delta d$ does not
significantly affect the width of the local light-cone at ${t,t+\delta t}$, and thereby limits
$\delta t$ from below. That is, the $t$-dependence  of the classical 3-metric $d(x,y)$ cannot vary too quickly in external time, so one can define a mean field or classical differential local light-cone. (b) or (c) could correspond to the non-commutative geometry scenario for nonlocality over the scale $\xi$ in the presence of a metric solution.
In the marginal case (d), where the scale of nonlocality is about that of the quantum fluctuations, the survival of the Lieb-Robinson local light-cone is too close to call.
Unfortunately, we really do not know which regime we live in. But if non-commutative geometry provided an action invariant under deformed diffeomorphisms  and nonlocal on scale $\xi$,
then that would be on-shell, and could naturally separate $\xi$ from quantum length fluctuations $\delta d$. \\

An alternative approach to relieve the classical vs. quantum tension inherent in $d(i,j)$ is to use semiclassical (coherent) states  $\Psi$ \cite{Thiemann_CohSt} of 3-space.
The idea is that each length on the initial slice $t$ has a quantum expectation $ \langle \hat{d} \rangle_{\Psi}\doteq d_{\Psi}$, a quantum fluctuation 
$\langle (\hat{d}-\langle\hat{d}\rangle_{\Psi})^{2}\rangle _{\Psi }\doteq \sigma^{2}_{d}(\Psi )$, and the classical length $d_{\mathrm{cl}}>0$.
It is also possible to use other geometric quantities besides $d(x,y)$, such as the areas of triangles or volumes of tetrahedra.
To achieve semiclassical consistency for $d(i,j)$ in a state $\Psi$, one would require for all sites $i,j$
\begin{align}
\frac{| d_{\mathrm{cl}} - d_{\Psi} | } {d_{\mathrm{cl}} } & \ll 1 \;\mathrm{and} \\
\frac{\sigma_{d} (\Psi )} {d_{\mathrm{cl}} } & \ll 1.
\end{align}
This makes the notion of classical distance insensitive to the quantum fluctuations from the state $\Psi$. Such a construction would also encounter difficulties
in case (d) above, where the range $\xi$ is quantum limited. \\

\section{Summary, Self-Criticism, and Conclusion}

In this work we have explored the consequences of non-ultralocal constraints within the context of the relational framework of canonical gravity.
It was shown that this leads to an on-shell non-Abelian algebra for the physical Hamiltonians, while the constraint algebra remains Abelian. Unitary propagators stay anomaly-free
 for smooth monotonic gauge flow in an external time parameter $t$.
A set of Hamiltonians that generate operator gauge-flow in $t$ with finite-ranged support patches was derived. After introducing a spatial
discretization, Lieb-Robinson bounds were reviewed and applied to demonstrate an on-shell differential time local light-cone. This local light-cone has the properties that there
is exponentially small norm leakage of discretized Dirac operator commutators outside the local light-cone, it displays suitable gauge (slicing) and (3+1)-diffeomorphism invariance, 
and the local light-cone can be ``integrated'' into  ``support tubes"  for discretized Dirac operators that resemble familiar causal curves from general relativity.
This entire Lieb-Robinson bound local light-cone structure collapses for ultralocal constraints.  
Therefore non-ultralocality together with Lieb-Robinson bounds go an unexpectedly long  way towards explaining how the standard quantum field theory version of 
micro-causality, where local observables commute at space-like distances, emerges from the (semi-) classical 
relational formulation of canonical gravity at length scales greater than that characterizing the nonlocality.  \\

Within the application of quantum field theory to fixed curved background spacetimes, one can derive the familiar causal advanced 
and retarded propagators as inverse
wave operators (Green's functions) for matter fields such as scalar bosons and so on. These show that the vacuum expectation values of commutators of 
canonical fields vanish outside the past or future light cone. However, while this is straightforward for Minkowski spacetime, to obtain unique solutions for general curved space times 
one imposes the stringent requirement that the manifold be globally hyperbolic. The Lieb-Robinson approach, on the other hand, requires no such corresponding {\it{ab initio}} 
strong global background causal structure assumption.\\

The criticism of the Lieb-Robinson bound route from non-ultralocality to local light-cones is abundantly clear from the responses to the questions in the previous section. 
While some issues, such as how two different classical spacetimes can
share the same Lieb-Robinson velocity, or the role of field gradients, are quite clear, many deeper concerns remain only partially clarified, or just display our glaring
ignorance.
These harder nuts to crack include: What are the origins of nonlocality?  
What is the detailed microscopic meaning of the range or correlation length $\xi=1/\mu$? What if $\xi$ is about the size of 
quantum fluctuations (marginal case)?
What are the specifics of the semiclassical limit
or choice of quantum states
necessary to ensure that quantum fluctuations do not destroy the local light-cone? Is it possible to handle the continuum limit more thoroughly than simply to assume the required limit is well-behaved? Might non-commutative geometry or field theory play a role in  these issues?
Each of these questions challenges us to probe more deeply into the ``atomic''  theory underlying the model of condensed matter ancestry presented here and
stands as motivation for future work.  \\

Nevertheless, it remains surprising that aspects of causality may be linked to non-ultralocality. Adopting ultralocality uncritically might be somewhat like what occurred in the 1950's with parity:
A beautiful symmetry, but Nature could be a lot more interesting if She broke it once in a while. \\

\section{Acknowledgements}
The author wishes to thank  B. Dittrich and S.T. Lu for helpful and insightful discussions, and P.D. Lu for his kind assistance in Shanghai and Putuo Shan where much of this work was carried out.\\

 \end{document}